\documentclass[12pt,english]{article}
\usepackage[T1]{fontenc}
\usepackage[latin9]{inputenc}
\usepackage{geometry}
\usepackage[active]{srcltx}
\usepackage{textcomp}
\usepackage{mathrsfs}
\usepackage{amsmath}
\usepackage{amssymb}
\usepackage{esint}

\makeatletter

\usepackage{mathdots}
\usepackage{amsfonts}
\usepackage{babel}
\usepackage{mathrsfs}
\usepackage{amsmath}
\usepackage{amssymb}
\usepackage{esint}
\usepackage{graphicx}
\usepackage{youngtab}

\makeatother

\usepackage{babel}

\begin{document}

\title{Higher Level String Resonances in Four Dimensions}

\author{Wan-Zhe Feng, Tomasz R. Taylor\\\emph{Department of Physics, Northeastern University, Boston, MA 02115, USA}}

\maketitle

\vspace{2cm}

\begin{abstract}
We study higher level Regge resonances of open superstrings, focusing on the universal part of the Neveu-Schwarz sector common to all D-brane realizations of the standard model. For Regge states with masses far above the fundamental string scale, we discuss the spin-dependence of their decay rates into massless gauge bosons. Extending our previous work on lowest level string excitations, we study the second mass level at which spins range from 0 to 3. We construct the respective vertex operators and compute the amplitudes involving one massive particle and two or three gauge bosons. To illustrate the use of BCFW recursion relations in superstring theory, we build the four-gluon amplitude from on-shell amplitudes involving string resonances and gauge bosons.
\end{abstract}

\vspace{6cm}

\section{Introduction and Summary}

Over the past quarter of century, research in superstring theory has been largely focused on
massless string excitations and their effective field theory description. Massless particles are the quantized string zero modes. There is a huge number of distinct massless spectra allowed by
various compactifications.
This arbitrariness limits the predictive power of the theory: it is referred to as the landscape problem. The quantization of first harmonic and higher level vibration modes yields massive particles --  Regge resonances with masses quantized in units of the fundamental mass scale $M$ as $\sqrt{n}M$ for $n$th harmonics. At the $n$th level, their spins  range from 0 to $n+1$.
The massive spectrum is also model-dependent, however it replicates massless states,
therefore for any string compactification that reproduces the standard model, there is a sharp prediction that quarks, gluons {\em etc}.\ will also appear among excited states.
The reason why there has not been much discussion of the properties of Regge resonances was the belief that the fundamental string scale must be of order of the Planck mass to explain the weakness of gravitational forces. This, however, has changed with the advent of D-brane constructions that allow arbitrary string scales because gravity can
``leak'' into large extra dimensions \cite{LED1}-\cite{LEDLMS2}. If the string scale is sufficiently low, excited gluons and other Regge resonances will be observed at the LHC
\cite{KK-1}-\cite{Carmi}.

In a recent paper \cite{1stSR}, we presented a detailed discussion of the ``universal'' part of the first massive level, common to all D-brane embeddings of the standard model. Here, we extend it to the second level, and discuss some general properties of higher levels. We are particularly interested in massive particles that couple to massless gauge bosons according to ``(anti)self-dual''  selection rules. These particles decay into two gauge bosons  with the same (say ++) helicities only and to more gluons in ``mostly plus'' helicity configurations.

We rely on the factorization techniques. They allow identifying not only the spins of Regge resonances propagating in a given channel, but also their couplings and decay rates. The paper is organized as follows. In Section 2 we perform the spin decomposition of the well-known four-gluon MHV amplitude in the $s$-channels of ($--$) and ($-$+) gluons. We examine decay rates of heavy states into two gluons, for masses much larger than $M$, {\em i.e}.\ in the large $n$ limit. We  find that for any particle with spin $j\leq n+1$, the maximum partial decay width into two gluons  is $n$-independent -- it never exceeds $M$. Particles with $j\sim \sqrt{n}=M_n/M$ have largest widths.
We also find that for $j\sim n$, the decay rate into two gluons is exponentially suppressed.
 In section 3, we study the second massive level. We construct the vertex operators for all ``universal'' bosons of the NS sector. We compute the amplitudes involving one such state and two or three gluons, focusing on the decays of the (anti)self-dual massive (complex) vector fields.

The amplitudes describing decays of heavy states into gauge bosons are also important for the superstring generalization of BCFW recursion relations to disk amplitudes with arbitrary number of external gauge bosons. Recently, it has been argued that the BCFW-deformed full-fledged string amplitudes have no singularities at the infinite value of the deformation parameter, therefore
BCFW recursion relations should be valid also in string theory \cite{StrBCFW1}-\cite{Fot3}. This approach to constructing the scattering amplitudes is however highly impractical because in order to increase the number of external massless particles from $N$ to $N+1$, one needs to compute an infinite number of amplitudes involving one massive state and $N-1$ massless ones, for all mass levels. It may be useful, however, for revealing some general properties of the amplitudes.
In section 4, we show that at least the four-gluon amplitude can be obtained by a BCFW deformation of a factorized sum involving on-shell amplitudes of one massive Regge state and two gauge bosons.

Although the main motivation for the present study is the exciting possibility of observing massive string states at the LHC, we should emphasize that elementary higher spin states can also exist outside the string context. Higher spin theory has been developed for quite a long time \cite{HSO1}-\cite{HSO8}. More recently, there is also a growing interest in the dynamics of higher spin states in string theory \cite{StrHS0}-\cite{StrHS8}. We hope that our study offers some new insight into the nature of higher spin interactions.

\section{Properties of Massive Superstring States Extracted by Factorizing  Four-Gluon Amplitudes}
The amplitudes describing the scattering of massless  superstring states (zero modes) encode many important properties of massive excitations. The spin content of intermediate massive particles, their decay rates {\em etc}.\ can be extracted by factorizing  massless amplitudes on their Regge poles \cite{Width}.
We are primarily interested in the properties of particles that couple to gauge bosons, {\em i.e}.\ of those that can be detected at particle accelerators if the fundamental string mass scale happens to be sufficiently low. As we will see below, even the simplest, four-gluon amplitudes contain some interesting information.

We will be using the helicity basis to describe gluon polarizations. For four gluons, all non-vanishing amplitudes can be obtained from a single, maximally helicity violating (MHV) configuration. Our starting point is
the well-known MHV amplitude \cite{GluAmp1,GluAmp2}
\begin{align}
\mathcal{M}(g_{1}^{-},g_{2}^{-},g_{3}^{+},g_{4}^{+})
& =4g^{2}\langle12\rangle^{4}\Big[\frac{V_{t}}{\langle12\rangle\langle23
\rangle\langle34\rangle\langle41\rangle}{\rm Tr}(T^{a_{1}}
T^{a_{2}}T^{a_{3}}T^{a_{4}}+T^{a_{2}}T^{a_{1}}T^{a_{4}}T^{a_{3}})\nonumber \\
 & \qquad\qquad+\frac{V_{u}}{\langle13\rangle\langle34
 \rangle\langle42\rangle\langle21\rangle}{\rm Tr}(T^{a_{2}}
 T^{a_{1}}T^{a_{3}}T^{a_{4}}+T^{a_{1}}T^{a_{2}}T^{a_{4}}T^{a_{3}})\nonumber \\
 & \qquad\qquad+\frac{V_{s}}{\langle14\rangle\langle42
 \rangle\langle23\rangle\langle31\rangle}{\rm Tr}(T^{a_{1}}
 T^{a_{3}}T^{a_{2}}T^{a_{4}}+T^{a_{3}}T^{a_{1}}T^{a_{4}}T^{a_{2}})\Big],
 \label{MHV--++}
\end{align}
where the Veneziano ``formfactor'' function reads

\begin{equation}
V_{t}=V(s,t,u)=\frac{\Gamma(1-s/M^{2})\Gamma(1-u/M^{2})}{\Gamma(1+t/M^{2})}.
\end{equation}
Here, $M^{2}=1/\alpha'$ is the fundamental string mass scale. $s,t,u$ are
the Mandelstam variables
\begin{equation} s=(k_1+k_2)^2~,\qquad t=(k_1+k_3)^2~,\qquad u=(k_2+k_3)^2\ .
\label{mandel}\end{equation}
while the spinor products $\langle\,\dots\rangle$ and $[\dots]$ are defined according to \cite{AmpEff}.
The momenta and helicities are specified for {\em incoming}\/ particles, therefore they need appropriate crossing to the relevant physical domains. In particular, $u<0$ and $t<0$ describing a $g_1g_2\to g_3g_4$ scattering process with $s>0$ can be expressed in terms of the scattering angle in the center of mass frame:
\begin{equation}
u=-\frac{s}{2}(1+\cos\theta),\qquad t=-\frac{s}{2}(1-\cos\theta),\label{mand1}
\end{equation}
so that $\theta=0$ describes forward scattering. Finally, $a_1,\dots, a_4$ are the gluon color indices. For future reference, it is convenient to absorb the gauge coupling $g$ into the color factors and define the combinations:
\begin{equation}
S^{a_1a_2}_{a_3a_4}=4g^2{\rm Tr}(\{T^{a_{1}}
T^{a_{2}}\}\{T^{a_{3}}T^{a_{4}}\})~,\qquad
A^{a_1a_2}_{a_3a_4}=4g^2{\rm Tr}([T^{a_{1}}
T^{a_{2}}][T^{a_{3}}T^{a_{4}}])\ .
\end{equation}
which are symmetric and antisymmetric, respectively, in the color indices of initial (and final) gluons.

Using the expansion in terms of s-channel resonances
\begin{equation}
B(-s/M^{2},-u/M^{2})=-\sum_{n=0}^{\infty}\frac{M^{2-2n}}{n!}
\frac{1}{s-nM^{2}}\left[\prod_{J=1}^{n}(u+M^{2}J)\right],\label{Bexpansion}
\end{equation}
we obtain, near the $n$th level pole ($s\rightarrow nM^{2}$),
\begin{equation}
V_{t}(n)=V(s,t,u)\approx\frac{1}{s-nM^{2}}\times
\frac{M^{2-2n}}{(n-1)!}\prod_{J=0}^{n-1}(u+M^{2}J).\label{Vexpansion}
\end{equation}
The spin content of Regge resonances can be disentangled by analyzing the angular distributions of scattered gluons, that is by decomposing the residue of each Regge pole in the basis of Wigner $d$-matrix elements $d_{m',m}^{(j)}(\theta)$.\footnote{Appendix A contains a brief introduction to Wigner d-matrices.}
In this context, $d_{m',m}^{(j)}(\theta)$ describe the angular distribution (in the center of mass frame) of the final gluons with the helicity difference $m=\lambda_3-\lambda_4$, produced in a decay of spin $j$ resonance;
$m'=\lambda_2-\lambda_1$ is the helicity difference of incident gluons \cite{Width}. Thus $m,~m'=0,\pm 2$.

We begin with the amplitude $\mathcal{M}(g_{1}^{-},g_{2}^{+},g_{3}^{+},g_{4}^{-})$
which
can be obtained from (\ref{MHV--++}) by interchanging $2\leftrightarrow4$. Near the lowest mass pole, associated to the "fundamental" $n=1$ string mode,
\begin{equation}
\mathcal{M}(g_{1}^{-},g_{2}^{+},g_{3}^{+},g_{4}^{-}) \xrightarrow{n=1}S^{a_1a_2}_{a_3a_4}\frac{M^{2}}{s-M^{2}}d_{2,2}^{(2)}\label{1st+-}
\end{equation}
which reflects the obvious fact \cite{1stSR} that in order to create two gluons with opposite helicities $(+ -)$ one needs a resonance with $j\ge 2$, which is the highest spin at this level. At the next $n=2$ level,
\begin{equation}
\mathcal{M}(g_{1}^{-},g_{2}^{+},g_{3}^{+},g_{4}^{-})  \xrightarrow{n=2}-A^{a_1a_2}_{a_3a_4}\frac{M^{2}}{s-2M^{2}}(\frac{2}{3})
(d_{2.2}^{(3)}+2d_{2,2}^{(2)}),\label{2nd+-}
\end{equation}
in agreement with \cite{Top}. Near the $n=3$ string resonance, we find,
\begin{equation}
\mathcal{M}(g_{1}^{-},g_{2}^{+},g_{3}^{+},g_{4}^{-})  \xrightarrow{n=3}S^{a_1a_2}_{a_3a_4}\frac{M^{2}}{s-3M^{2}}\frac{3}{56}
(9d_{2,2}^{(4)}+21d_{2,2}^{(3)}+26d_{2,2}^{(2)})\label{3rd+-}
\end{equation}
In general, at the $n$th massive level, states with all spins from
2 up to $n+1$ appear in the $s$-channel, decaying into two opposite helicity gluons.
The residues of Regge poles factorize as
\begin{equation}
\makebox{Res}_{s=nM^2}\mathcal{M}(g_{1}^{-},g_{2}^{+},g_{3}^{+},g_{4}^{-})=\sum_{j=2}^{n+1}\sum_a
F^{aj}_{+-;a_1a_2}(F^{aj}_{+-;a_3a_4})^*d_{2,2}^{(j)}(\theta)\label{fact+-}
\end{equation}
where $F$ are the matrix elements for the decay of a spin $j$ resonance, in the $m_j=2$ eigenstate (in the the center of mass frame), into two gluons moving along the $\pm z$ axis, with helicities $\pm 1$, respectively \cite{Width}. In the above expression, the sum over intermediate color indices appears after rewriting the color factors as
\begin{eqnarray}
S^{a_1a_2}_{a_3a_4}&=&\sum_a(4\sqrt{2}gd^{a_1a_2 a})(4\sqrt{2}gd^{a_3a_4 a})\\
-A^{a_1a_2}_{a_3a_4}&=&\sum_a(\sqrt{2}gf^{a_1a_2 a})(\sqrt{2}gf^{a_3a_4 a})
\end{eqnarray}
where $f$ are the gauge group structure constants while $d$ are the symmetrized traces:
\begin{equation}
d^{a_1a_2a_3}=\makebox{STr}(T^{a_1}T^{a_2}T^{a_3})\ .
\end{equation}
The matrix elements involve totally symmetric group factors at odd levels and antisymmetric ones at even levels. This can be understood as a consequence of world-sheet parity \cite{Width,1stSR}. Note that the numerical factors multiplying $d$-functions in
Eqs.(\ref{1st+-})-(\ref{3rd+-}) and at higher $n$  are positive, as required by unitarity, {\em c.f}.\ Eq.(\ref{fact+-}).

The amplitude $\mathcal{M}(g_{1}^{-},g_{2}^{+},g_{3}^{-},g_{4}^{+})$
can be obtained from (\ref{MHV--++}) by interchanging $2\leftrightarrow 3$, however there is no need to repeat calculations because it can be also obtained from $\mathcal{M}(g_{1}^{-},g_{2}^{+},g_{3}^{+},g_{4}^{-})$ by interchanging the color indices $a_3\leftrightarrow a_4$ combined with the reflection $\theta\to\pi-\theta$, for which
\[
\left\{ \begin{array}{ll}
d_{2,2}^{(j)}(-\cos\theta)=(-1)d_{2,-2}^{(j)}(\cos\theta) & \qquad j\; \makebox{odd}\\
d_{2,2}^{(j)}(-\cos\theta)=d_{2,-2}^{(j)}(\cos\theta) & \qquad j\; \makebox{even}
\end{array}\right.
\]
As a result, $d_{2,2}^{(j)}\to d_{2,-2}^{(j)}$ and the coefficients acquire alternating $(-1)^{n+j+1}$ signs, for instance
\begin{equation}
\mathcal{M}(g_{1}^{-},g_{2}^{+},g_{3}^{-},g_{4}^{+}) \xrightarrow{n=3}S^{a_1a_2}_{a_3a_4}\frac{M^{2}}{s-3M^{2}}
\frac{3}{56}(9d_{2,-2}^{(4)}-21d_{2,-2}^{(3)}+26d_{2,-2}^{(2)}).
\end{equation}

Next,  we turn to the amplitude $\mathcal{M}(g_{1}^{-},g_{2}^{-},g_{3}^{+},g_{4}^{+})$.
This case is very interesting because the resonances appearing in the $s$-channel couple only to (anti)self-dual gauge field configurations, {\em i.e}.\ to gluons in $(++)$ or $(--)$ helicity configurations.
In the previous work \cite{1stSR}, we discussed the first massive
level and identified a complex scalar $\Phi$ (2 degrees of freedom $\Phi\equiv\Phi_{+}$ and $\bar{\Phi}\equiv\Phi_{-}$)
which couples to gluons according to the selection rules
\begin{equation}
\mathscr{A}\left[\Phi_{+},-,-\right]=\mathscr{A}
\left[\Phi_{+},+,-\right]=\mathscr{A}\left[\Phi_{-},+,+\right]=
\mathscr{A}\left[\Phi_{-},+,-\right]=0.\label{1stScalarSR}
\end{equation}
This scalar is the sole resonance contributing to
\begin{equation}
\mathcal{M}(g_{1}^{-},g_{2}^{-},g_{3}^{+},g_{4}^{+})\xrightarrow{n=1}S^{a_1a_2}_{a_3a_4}
\frac{M^{2}}{s-M^{2}}d_{0,0}^{(0)}\ .
\end{equation}
At higher levels, there are more such particles, with higher spins:
\begin{align}
\mathcal{M}(g_{1}^{-},g_{2}^{-},g_{3}^{+},g_{4}^{+}) & \xrightarrow{n=2}-A^{a_1a_2}_{a_3a_4} \frac{2M^{2}}{s-2M^{2}}(d_{0,0}^{(1)}),\label{2ndVector}\\
\mathcal{M}(g_{1}^{-},g_{2}^{-},g_{3}^{+},g_{4}^{+}) & \xrightarrow{n=3}S^{a_1a_2}_{a_3a_4} \frac{3M^{2}}{s-3M^{2}}(\frac{3}{4}d_{0,0}^{(2)}+\frac{1}{4}d_{0,0}^{(0)}),\\
\mathcal{M}(g_{1}^{-},g_{2}^{-},g_{3}^{+},g_{4}^{+}) & \xrightarrow{n=4}-A^{a_1a_2}_{a_3a_4} \frac{4M^{2}}{s-4M^{2}}(\frac{8}{15}d_{0,0}^{(3)}+\frac{7}{15}d_{0,0}^{(1)}),\\
\mathcal{M}(g_{1}^{-},g_{2}^{-},g_{3}^{+},g_{4}^{+}) & \xrightarrow{n=5}S^{a_1a_2}_{a_3a_4} \frac{5M^{2}}{s-5M^{2}}(\frac{125}{336}d_{0,0}^{(4)}+\frac{125}{252}
d_{0,0}^{(2)}+\frac{19}{144}d_{0,0}^{(0)}).
\end{align}

In order to proceed to higher $n$, we first note that, in this case,
\begin{equation}
d_{0,0}^{(l)}(\theta)=P_l(\cos\theta),
\end{equation}
see Appendix A, therefore the resonance coefficients can be obtained by decomposing the angular dependence in the basis of Legendre polynomials:
\begin{align}
\mathcal{M}(g_{1}^{-},g_{2}^{-},g_{3}^{+},g_{4}^{+}) & \xrightarrow{{\rm odd}~ n}S^{a_1a_2}_{a_3a_4}\frac{M^{2}}{s-nM^{2}}\sum_{k=0,2\cdots}^{n-1}c_{k}^{(n)}P_{k}
(\cos\theta)\ ,\label{decomp-odd}\\
\mathcal{M}(g_{1}^{-},g_{2}^{-},g_{3}^{+},g_{4}^{+}) & \xrightarrow{{\rm even}~n}-A^{a_1a_2}_{a_3a_4}\frac{M^{2}}{s-nM^{2}}\sum_{k=1,3\cdots}^{n-1}c_{k}^{(n)}P_{k}
(\cos\theta)\ .\label{decomp-even}
\end{align}
The above expansions involve even Legendre polynomials only for odd $n$ and odd ones for even $n$,
reflecting the $g_3\leftrightarrow g_4$ ($a_3\leftrightarrow a_4,~ \theta\to
\pi-\theta$) symmetry of the amplitude.
A straightforward, but tedious computation, outlined
in Appendix B, yields the following coefficients:
\begin{align}
c_{k}^{(n)} & =\frac{n}{(n-1)!}\sum_{j=0}^{\frac{n-1-k}{2}}\sum_{i=0}^{(n-1-k-2j)}\frac{(-1)^{n-1-k-2j}}{2^{2j+i-1}}\frac{(2k+1)(k+j+1)!(k+2j+i)!}{i!j!(2k+2j+2)!}\nonumber \\
 & \qquad\qquad\qquad\qquad\qquad\qquad\times(n)^{k+2j}(n-2)^{i}s(n-1,k+2j+i),\label{coeff}
\end{align}
where $s(n,k)$ is the Stirling number of the first kind, defined through the expansion of the Pochhammer symbol:
\begin{equation}
(x)_{n}=\frac{\Gamma(x+n)}{\Gamma(x)}=x(x+1)...(x+n-1)=
\sum_{k=0}^{n}(-1)^{n-k}s(n,k)x^{k}.\label{Stirling1}
\end{equation}

We want to see how the decay rates of Regge resonances at a given mass level $n$ depend on their spin $j$ and in general, on the $n,j$ dependence of their partial widths into two gluons, in the large $n$ limit. It has been often suggested that string perturbation theory breaks down at energies much higher than the fundamental string mass, with the onset of non-perturbative effects marked by large widths of Regge particles, covering up the mass gap between subsequent resonances. To that end, we examine the $k$-dependence of the coefficients $c_{k}^{(n)}$, see Eq.(\ref{coeff}), in the large $n$ limit. Since we could not find a compact expression for Stirling numbers, we had to resort to numerical methods. On Figure 1, we plot $c_{k}^{(n)}$ as a function of $k$ for two typical values, $n=32^2$ and $n=50^2$. For small $k$, roughly $k\sim \sqrt{n}$, one finds a sharp peak at $k\approx 2\sqrt{n}$, with $c_{2\sqrt{n}}^{(n)}\approx \sqrt{n}/2$. For large $k\sim n$, the coefficients are exponentially suppressed. For example,
\begin{equation}
c_{n-1}^{(n)}=\frac{n^{n}(n-1)!}{(2n-2)!} ~\xrightarrow{{\rm large}~n}~ \Big(\frac{2n}{\sqrt{e}}\Big)^{-2n}\ ,
\end{equation}
see Appendix B. Since $\sum_{j~\rm odd}^{n-1}c_{j}^{(n)}=\sum_{j~\rm even}^{n-1}c_{j}^{(n)}=n$, we conclude that the sums are saturated by spins ranging from 0 to $j\sim\sqrt{n}$, with the maximum
$c_{\rm max}\sim\sqrt{n}$.
\begin{figure}
\centering
\includegraphics[width=8cm]{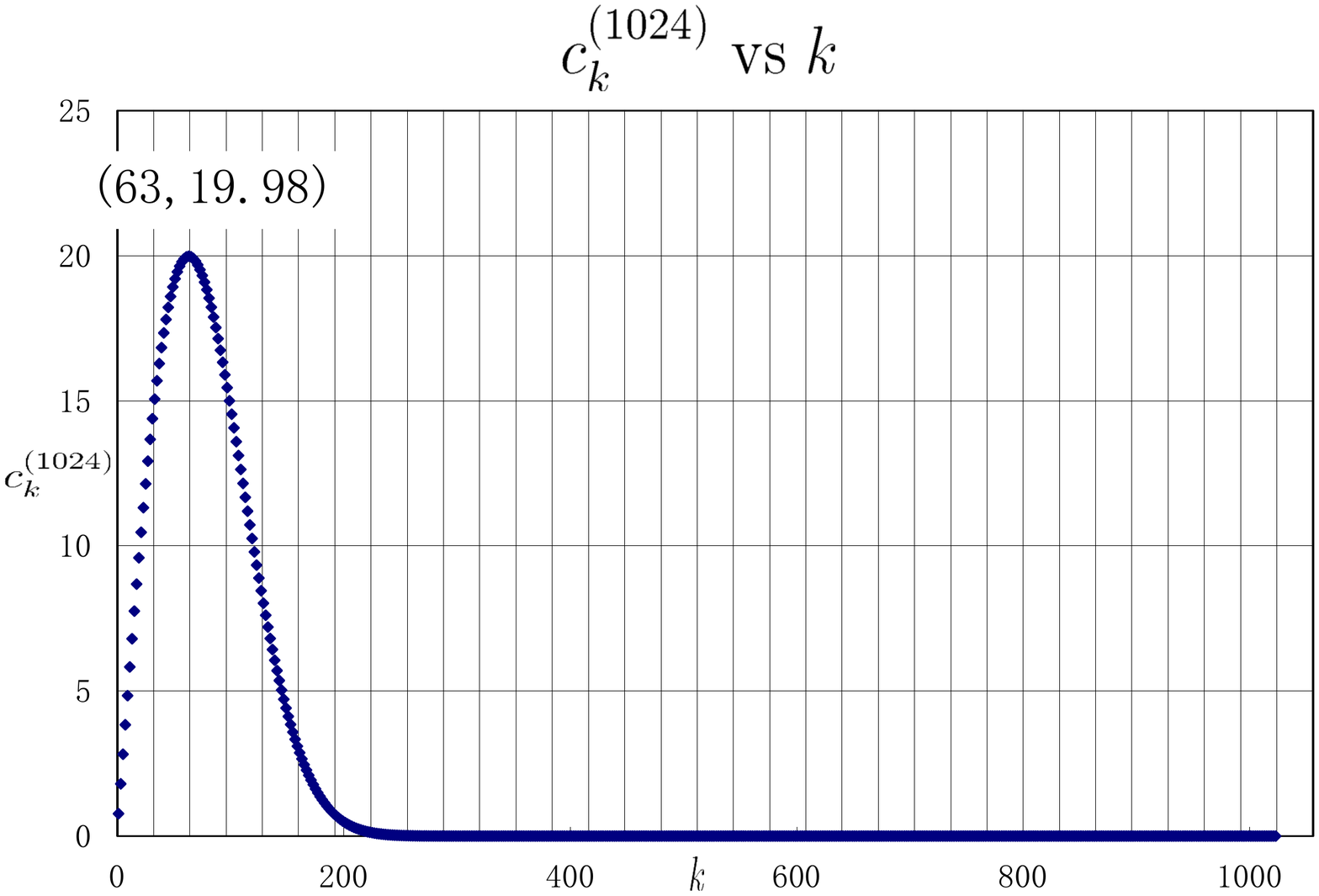}
\includegraphics[width=8cm]{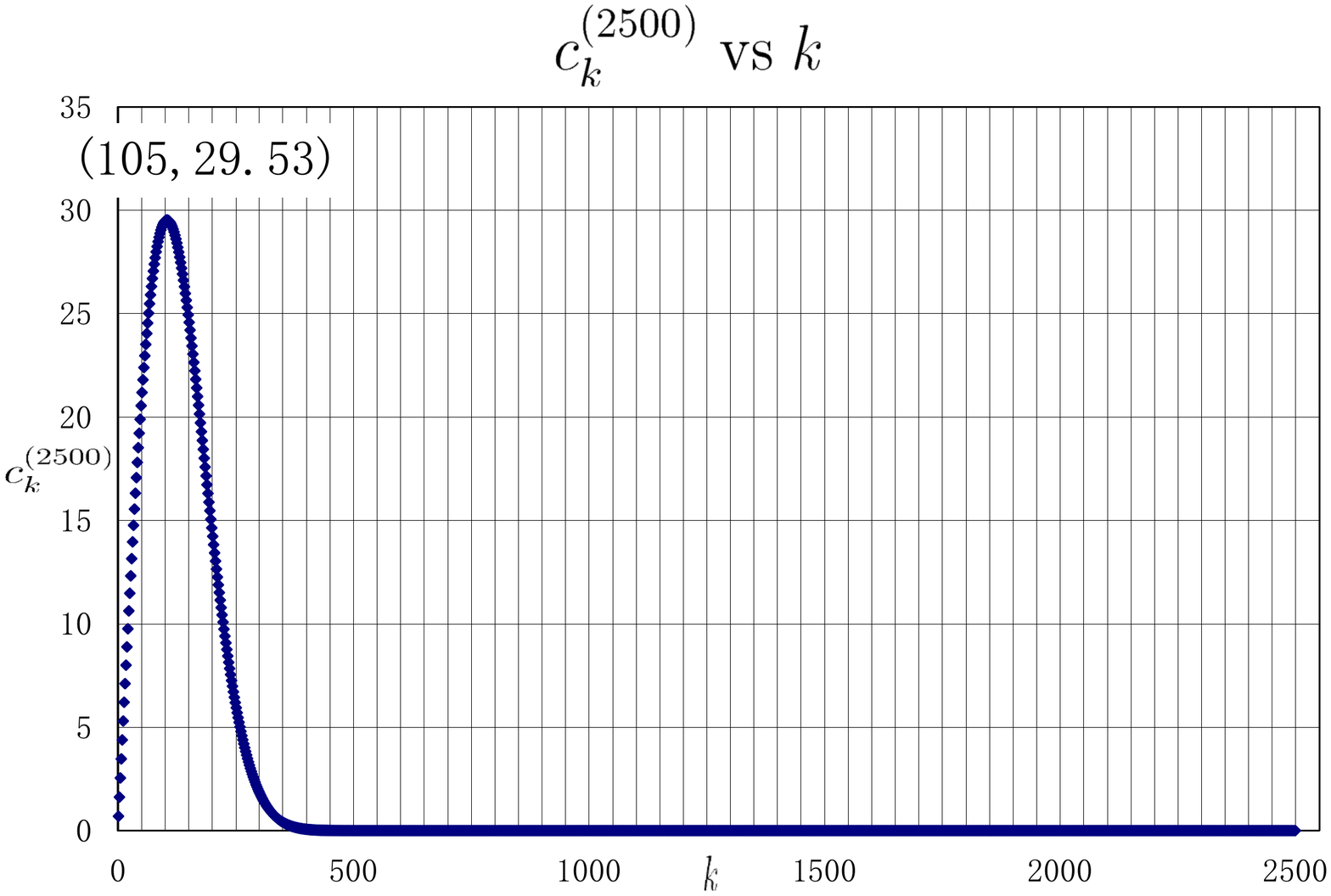}
\caption{\label{fig.1} The coefficients of $P_k(\cos \theta)$ at the 1024th and 2500th massive levels. On the x-axis, we mark multiples of $\sqrt{n}$ to display the peaks at $k\approx 2\sqrt{n}$ }
\end{figure}

The partial width of mass $M_n=\sqrt{n}M$, spin $j$ resonance $R_{n,j}$ into a pair of gluons is given by \cite{Width}
\begin{equation}
\Gamma(R_{n,j}\to gg)=g^2\delta\,\frac{c_{j}^{(n)}M^2}{32(2j+1)\pi M_n}
\end{equation}
where $\delta\sim 1$ is the gauge group factor. In the denominator, the number $2j+1$ comes from averaging over spin components, and provides additional suppression for large $j$, however we will not take it into account because it is purely statistical.
Thus the width size is determined by the ratio $c_{j}^{(n)}M^2/M_n$. From our discussion of the coefficients, it follows that the largest possible widths are $n$-independent,  $(2j+1)\Gamma(n\to\infty,j\sim \sqrt{n})\sim M$, the same as for low-lying Regge resonances. We conclude that the leading order (disk) approximation gives a perfectly sensible result for the decays of higher level Regge resonances. Note that the exponential suppression of direct decays of very high spin ($j\sim n$) particles into two massless gluons is akin to the Sudakov formfactor. These particles will cascade into lower mass, lower spin states, decaying at the end into a large number of gluons.

\section{The Second Massive Level: Physical States, Vertices and Amplitudes}
We will be using the \emph{Old
Covariant Quantization (OCQ)} method for identifying the physical states.
The second massive level has been previously discussed in Ref.\cite{Het2nd}, in the context of ten-dimensional heterotic superstrings. Here, we focus on four-dimensional open string excitations, especially on those that can be created by the fusion of gauge bosons associated to strings ending on D-branes. Such particles appear in the Neveu-Schwarz (NS) sector and are universal to the whole landscape of models because their vertices do not contain internal parts associated to compact dimensions.
As a warm-up, we start from the first massive level, worked out in Ref.\cite{1stSR},
using it as a check of the method.

\subsection{The First Massive Level}

In the NS sector, the four-dimensional string states are created by $SO(3,1)$ Lorentz-covariant creation operators acting on the vacuum. At the first massive level, their numbers must add up to $-$3/2, therefore the states can be written as
\begin{equation}
|n=1\rangle=\Big(\chi_{1\mu}\psi_{-\frac{3}{2}}^{\mu}+\chi_{2\mu\nu}\alpha_{-1}^{\mu}\psi_{-\frac{1}{2}}^{\nu}+\chi_{3\mu\nu\rho}\psi_{-\frac{1}{2}}^{\mu}\psi_{-\frac{1}{2}}^{\nu}\psi_{-\frac{1}{2}}^{\rho}\Big)|0;k\rangle,\label{GE1st}
\end{equation}
where $|0;k\rangle$ is the open string vacuum state in the NS sector. Here, the Greek letters denote $D=4$ spacetime indices. Note that
$\chi_{3\mu\nu\rho}$ is totally antisymmetric due to anticommuting $\psi$ operators.
The physical state conditions are:
\begin{equation}
(L_{0}-\frac{1}{2})|n=1\rangle=0,\qquad L_{1}|n=1\rangle=0,\qquad G_{\frac{3}{2}}|n=1\rangle=G_{\frac{1}{2}}|n=1\rangle=0,\label{PSC1st}
\end{equation}
where the superconformal Virasoro generators read,
\begin{eqnarray}
L_{m} & = & \frac{1}{2}\sum_{n}:\alpha_{m-n}^{\lambda}\alpha_{n\lambda}:+\frac{1}{4}\sum_{r}(2r-m):\psi_{m-r}^{\lambda}\psi_{r\lambda}:+a\delta_{m,0},\label{SVG-L}\\
G_{r} & = & \sum_{n}\alpha_{n}^{\lambda}\psi_{(r-n)\lambda},\label{SVG-G}
\end{eqnarray}
and $a=0$ in the NS sector. The first condition in (\ref{PSC1st})
gives the mass shell condition $k^{2}=1/\alpha'=M^2$ for the first massive
level, as expected. By using the commutation relations of the bosonic
and fermionic operators: $[\alpha_{m}^{\mu}\alpha_{n}^{\nu}]=m\eta^{\mu\nu}\delta_{m,-n}$,
$\{\psi_{r}^{\mu},\psi_{s}^{\nu}\}=\eta^{\mu\nu}\delta_{r,-s}$,
the three remaining conditions of
Eqs.(\ref{PSC1st}) yield:

\begin{align}
\sqrt{2\alpha'}\chi_{1\mu}k^{\mu}+\chi_{2\mu\nu}\eta^{\mu\nu} & =0,\label{PSC1st-1}\\
\chi_{1\mu}+\sqrt{2\alpha'}\chi_{2\mu\nu}k^{\nu} & =0,\label{PSC1st-2}\\
\chi_{2\mu\nu}-\chi_{2\nu\mu}+6\sqrt{2\alpha'}\chi_{3\mu\nu\rho}k^{\rho} & =0.\label{PSC1st-3}
\end{align}
In order to simplify the above constraints, it is convenient to decompose
\begin{equation}
\chi_{2\mu\nu}=S_{2(\mu\nu)}+A_{2[\mu\nu]},
\end{equation}
where $S_{2(\mu\nu)}$ and $A_{2[\mu\nu]}$ are the symmetric and
antisymmetric parts of $\chi_{2\mu\nu}$ respectively.
Then the symmetric and antisymmetric parts decouple in (\ref{PSC1st-1})-(\ref{PSC1st-3}). The symmetric one is subject to
\begin{equation}
\left\{ \begin{aligned} & \sqrt{2\alpha'}\chi_{1\mu}k^{\mu}+S_{2(\mu\nu)}\eta^{\mu\nu}=0\\
 & \chi_{1\mu}+\sqrt{2\alpha'}S_{2(\mu\nu)}k^{\nu}=0
\end{aligned}
\right.,\label{1stEQS-1}
\end{equation}
which is fairly easy to resolve. We obtain the following solutions:
\begin{enumerate}
\item $S_{2(\mu\nu)}=\alpha_{\mu\nu}$ and $\chi_{1\mu}=0$, where $\alpha_{\mu\nu}$
is a spin-2 field satisfying $\alpha_{\mu\nu}k^{\nu}=\alpha_{\mu\nu}\eta^{\mu\nu}=0$.
\item $S_{2(\mu\nu)}=\sqrt{2\alpha'}(k_{\mu}\xi_{\nu}+\xi_{\mu}k_{\nu})$
and $\chi_{1\mu}=2\xi_{\mu}$, where $\xi_{\mu}$ represents a spin-1
field satisfying $\xi_{\mu}k^{\mu}=0$.
\item $S_{2(\mu\nu)}=\eta_{\mu\nu}+2\alpha'k_{\mu}k_{\nu}$ and $\chi_{1\mu}=\sqrt{2\alpha'}k_{\mu}$.
\end{enumerate}
In this way, we obtain a spin-2 field, a vector field and a scalar field. At this point,
let us count the physical degrees of freedom to make sure we are not losing any
states. We started from one symmetric Lorentz 2-tensor $S_{2(\mu\nu)}$
which has 10 d.o.f.\ and one Lorentz vector $\chi_{1\mu}$ which has
4 d.o.f. On the other hand,  Eqs.(\ref{1stEQS-1}) gave us $1+4=5$ constraints. Thus
we are left with $14-5=9$ d.o.f., which are exactly the degrees of
freedom of a spin-2 field (5 d.o.f.), a vector field (3 d.o.f.) and
a scalar (1 d.o.f.).

The antisymmetric part of $\chi_{2\mu\nu}$ is also easy to handle. Eqs.(\ref{PSC1st-1})-(\ref{PSC1st-3}) boil down to
\begin{equation}
A_{2[\mu\nu]}+3\sqrt{2\alpha'}\chi_{3\mu\nu\rho}k^{\rho}=0.\label{1stEQS-2}
\end{equation}
The solutions are:
\begin{enumerate}
\item $\chi_{3\mu\nu\rho}=i\varepsilon_{\mu\nu\rho\sigma}k^{\sigma}$, $A_{2[\mu\nu]}=0$,
and $\varepsilon_{\mu\nu\rho\sigma}$ is the Levi-Civita symbol.
\item $\chi_{3\mu\nu\rho}=\varepsilon_{\mu\nu\rho\sigma}\xi'^{\sigma}$
and $A_{2[\mu\nu]}=-3\varepsilon_{\mu\nu\rho\sigma}k^{\rho}\xi'^{\sigma}$.
$\xi'_{\mu}$ is another spin-1 field satisfying $\xi'_{\mu}k^{\mu}=0$.
\end{enumerate}
In this way, we obtain a pseudo-vector (3 d.o.f.) and a pseudo-scalar (1 d.o.f.).
To recapitulate, we started
from a 3-form $\chi_{3\mu\nu\rho}$ (4 d.o.f.) and an antisymmetric
2-tensor $A_{2[\mu\nu]}$ (6 d.o.f.). Eq.(\ref{1stEQS-2}) gave us 6
constraints. Thus we are left with $10-6=4$ d.o.f., which are exactly
what we get.

In order to construct the vertex operators, we use the state-operator correspondence and replace
the bosonic and fermionic creation operators with world-sheet bosons
and fermions as follows:
\begin{align}
\alpha_{-m}^{\mu} & \rightarrow i\sqrt{\frac{1}{2\alpha'}}\frac{1}{(m-1)!}\partial^{m}X^{\mu},\\
\psi_{-r}^{\mu} & \rightarrow\frac{1}{(r-\frac{1}{2})!}\partial^{r-\frac{1}{2}}\psi^{\mu}.
\end{align}
Therefore,  we have the following vertices, universal to all $D=4$ compactifications,
which satisfy the physical state conditions. They are: a spin-2 field,
\begin{equation}
V_{\alpha}=\alpha_{\mu\nu}\sqrt{\frac{1}{2\alpha'}}i\partial X^{\mu}\psi^{\nu}e^{-\phi}e^{ikX},
\end{equation}
with $\alpha_{\mu\nu}k^{\nu}=\alpha_{\mu\nu}\eta^{\mu\nu}=0$; one
spin-1 field and one pseudo spin-1 field,
\begin{align}
V_{\xi} & =(\xi_{\mu}k_{\nu}+k_{\mu}\xi_{\nu})i\partial X^{\mu}\psi^{\nu}e^{-\phi}e^{ikX}+2\xi_{\mu}\partial\psi^{\mu},\label{1stNv1}\\
V_{\xi'} & =\varepsilon_{\mu\nu\rho\sigma}\xi'^{\sigma}\psi^{\mu}\psi^{\nu}\psi^{\rho}e^{-\phi}e^{ikX}-3\varepsilon_{\mu\nu\rho\sigma}k^{\rho}\xi'^{\sigma}i\partial X^{\mu}\psi^{\nu}e^{-\phi}e^{ikX},\label{1stNv2}
\end{align}
with $\xi_{\mu}k^{\mu}=\xi'_{\mu}k^{\mu}=0$; plus one scalar and
one pseudo-scalar,
\begin{align}
V_{ps} &  \label{phi1} =\sqrt{2\alpha'}i\varepsilon_{\mu\nu\rho\sigma}k^{\sigma}\psi^{\mu}\psi^{\nu}\psi^{\rho}e^{-\phi}e^{ikX},\\
V_{s} & =\Big[(\eta_{\mu\nu}+2\alpha'k_{\mu}k_{\nu})\sqrt{\frac{1}{2\alpha'}}i\partial X^{\mu}\psi^{\nu}+\sqrt{2\alpha'}k_{\mu}\partial\psi^{\mu}\Big]e^{-\phi}e^{ikX}.\label{phi2}
\end{align}

It is well known that not all fields satisfying the physical state conditions
like (\ref{PSC1st}) appear in the spectrum. Actually,
both spin-1 vertices (\ref{1stNv1}) and
(\ref{1stNv2}) represent such null, spurious states, decoupled from the rest of the spectrum.\footnote{A spurious
state is defined to be a state that is orthogonal to all
the physical states, and a null state is defined to be a spurious
state that satisfies the physical state conditions \cite{StringPol}.%
}
This can be demonstrated by computing their two-point correlation functions and showing that they do not contain poles appropriate to physical propagators. It is also easy to show that they do not couple to two gauge bosons in any helicity configuration: the three-point amplitude involving two gauge bosons and one such massive state is zero.

To summarize, at the first massive level of NS sector, we have a total of 7 universal degrees
of freedom.
They are a spin-2 field $\alpha_{\mu\nu}$, plus scalar and a pseudoscalar. As explained in
Ref.\cite{1stSR}, it is natural to combine Eqs.(\ref{phi1}) and (\ref{phi2}) into one vertex of a ``self-dual'' complex scalar,
\begin{equation}
V_{\Phi_{\pm}}=\Big[(\eta_{\mu\nu}+2\alpha'k_{\mu}k_{\nu})\sqrt{\frac{1}{2\alpha'}}i\partial X^{\mu}\psi^{\nu}+\sqrt{2\alpha'}k_{\mu}\partial\psi^{\mu}\pm\frac{i}{6}\sqrt{2\alpha'}
\varepsilon_{\mu\nu\rho\sigma}k^{\sigma}\psi^{\mu}\psi^{\nu}\psi^{\rho}\Big]e^{-\phi}e^{ikX}\ ,
\end{equation}
which satisfies the selection rules written in Eq.(\ref{1stScalarSR}). We will find similar complex vector resonances at the second level.

\subsection{The Second  Level}

At the second level, the number of creation operators add up to $-5/2$:
\begin{align}
|n=2\rangle & =\Big(\zeta_{1\mu}\psi_{-\frac{5}{2}}^{\mu}+\zeta_{2\mu\nu}\alpha_{-1}^{\mu}
\psi_{-\frac{3}{2}}^{\nu}+\zeta'_{2\mu\nu}\alpha_{-2}^{\mu}\psi_{-\frac{1}{2}}^{\nu}+
\zeta_{3\mu\nu\rho}\alpha_{-1}^{\mu}\alpha_{-1}^{\nu}\psi_{-\frac{1}{2}}^{\rho}+
\zeta'_{3\mu\nu\rho}\psi_{-\frac{1}{2}}^{\mu}\psi_{-\frac{1}{2}}^{\nu}
\psi_{-\frac{3}{2}}^{\rho}\nonumber \\
 & \qquad+\zeta_{4\mu\nu\rho\sigma}\alpha_{-1}^{\mu}
 \psi_{-\frac{1}{2}}^{\nu}\psi_{-\frac{1}{2}}^{\rho}
 \psi_{-\frac{1}{2}}^{\sigma}+\zeta_{5\mu\nu\rho
 \sigma\gamma}\psi_{-\frac{1}{2}}^{\mu}\psi_{-\frac{1}{2}}^{\nu}
 \psi_{-\frac{1}{2}}^{\rho}\psi_{-\frac{1}{2}}^{\sigma}
 \psi_{-\frac{1}{2}}^{\gamma}\Big)|0;k\rangle.\label{GE2nd}
\end{align}
The physical state conditions are:
\begin{enumerate}
\item $(L_{0}-\frac{1}{2})|n=2\rangle=0$,
\item $L_{2}|n=2\rangle=L_{1}|n=2\rangle=0$,
\item $G_{\frac{5}{2}}|n=2\rangle=G_{\frac{3}{2}}|n=2\rangle=G_{\frac{1}{2}}|n=2\rangle=0$,
\end{enumerate}
with the superconformal Virasoro generators written in (\ref{SVG-L}) and (\ref{SVG-G}).
Here again, the first condition amounts to
$k^{2}=2/\alpha'=2M^2$.
To solve the remaining constraints, it is convenient to decompose the tensors, especially those of higher rank, into representations that are symmetric or antisymmetric in groups of Lorentz indices. This is most succinctly done by using Young tableaux. Our analysis parallels the discussion of the heterotic case (in ten dimensions) presented in Ref.\cite{Het2nd}.
The tensors $\zeta_{2\mu\nu}$ and $\zeta'_{2\mu\nu}$
can be decomposed into  symmetric and antisymmetric parts:
\begin{equation}
\zeta_{2\mu\nu}=S_{2(\mu\nu)}+A_{2[\mu\nu]},\qquad\zeta'_{2\mu\nu}=S'_{2(\mu\nu)}+A'_{2[\mu\nu]}.
\end{equation}
The rank 3 tensors $\zeta_{3\mu\nu\rho}$ and $\zeta'_{3\mu\nu\rho}$ can be
decomposed as
\begin{align}
\zeta_{3\mu\nu\rho} & \rightarrow S_{3(\mu\nu\rho)}+B_{3(\mu[\nu)\rho]}+D_{3[\mu(\nu]\rho)}+A_{3[\mu\nu\rho]},\\
\zeta'_{3\mu\nu\rho} & \rightarrow S'_{3(\mu\nu\rho)}+B'_{3(\mu[\nu)\rho]}+D'_{3[\mu(\nu]\rho)}+A'_{3[\mu\nu\rho]},
\end{align}
corresponding to
\begin{gather}
\Yvcentermath1\young(\mu)\otimes\young(\nu)\otimes\young(\rho)=\young(\mu\nu\rho)\oplus\young(\mu\nu,\rho)\oplus\young(\mu\rho,\nu)\oplus\young(\mu,\nu,\rho)\;,
\end{gather}
or by dimensions,
\begin{equation}
\text{{\bf 4}\;\ensuremath{\otimes}\;{\bf 4}\;\ensuremath{\otimes}\;{\bf 4}\;=\;{\bf 20}\;\ensuremath{\oplus}\;{\bf 20}\;\ensuremath{\oplus}\;{\bf 20}\;\ensuremath{\oplus}\;{\bf 4}}\;.
\end{equation}
Due to the (anti)commutation properties of the creation operators in  (\ref{GE2nd}), we can set $D_{3[\mu(\nu]\rho)}=A_{3[\mu\nu\rho]}=S'_{3(\mu\nu\rho)}=B'_{3(\mu[\nu)\rho]}=0$. We are left with
\begin{align}
\zeta_{3\mu\nu\rho} & =S_{3(\mu\nu\rho)}+B_{3(\mu[\nu)\rho]},\qquad\zeta'_{3\mu\nu\rho}=D'_{3[\mu(\nu]\rho)}+A'_{3[\mu\nu\rho]}.
\end{align}
Similarly, the rank 4 tensor $\zeta_{4\mu\nu\rho\sigma}$ can be decomposed
as
\begin{gather}
\Yvcentermath1\young(\mu)\otimes\young(\nu)\otimes\young(\rho)\otimes\young(\sigma)=\young(\mu\nu\rho\sigma)\oplus\young(\mu\nu\rho,\sigma)\oplus\young(\mu\nu\sigma,\rho)\oplus\young(\mu\rho\sigma,\nu)\oplus\young(\mu\nu,\rho\sigma)\oplus\young(\mu\rho,\nu\sigma)\nonumber \\
\Yvcentermath1\oplus\;\young(\mu\nu,\rho,\sigma)\oplus\young(\mu\rho,\nu,\sigma)\oplus\young(\mu\sigma,\nu,\rho)\oplus\young(\mu,\nu,\rho,\sigma)\;,
\end{gather}
or by dimensions,
\begin{equation}
\text{{\bf 4}\;\ensuremath{\otimes}\;{\bf 4}\;\ensuremath{\otimes}\;{\bf 4}\;\ensuremath{\otimes}\;{\bf 4}\;=\;{\bf 35}\;\ensuremath{\oplus}\;{\bf 45}\;\ensuremath{\times}\;3\;\ensuremath{\oplus}\;{\bf 20}\;\ensuremath{\times}\;2\;\ensuremath{\oplus}\;{\bf 15}\;\ensuremath{\times}\;3\;\ensuremath{\oplus}\;{\bf 1}}\;.
\end{equation}
Here again, we can ignore all but the
last four Young diagrams. Actually, due to the anticommutation of $\psi$ operators in the respective term of (\ref{GE2nd}), the three 3-row diagrams
would lead to the same state, therefore we are allowed to pick just one of
of them, say the  one symmetric in $\mu$ and $\nu$.
Thus the 4-tensor is decomposed as
\begin{equation}
\zeta_{4\mu\nu\rho\sigma}\rightarrow B_{4(\mu[\nu)\rho\sigma]}+A_{4[\mu\nu\rho\sigma]}.
\end{equation}
Finally, the term involving completely antisymmetric $\zeta_{5\mu\nu\rho\sigma\gamma}$ must necessarily involve one internal index, therefore we do not discuss it any further.

The second physical state condition, $L_{2}|n=2\rangle=L_{1}|n=2\rangle=0$,
yields,
\begin{align}
2\zeta_{1\mu}+\sqrt{2\alpha'}(S_{2(\mu\nu)}-A_{2[\mu\nu]})k^{\nu} & =0,\label{PSC-1}\\
A'_{3[\mu\nu\rho]}+\sqrt{2\alpha'}(B_{4(\sigma[\mu)\nu\rho]}+A_{4[\sigma\mu\nu\rho]})k^{\sigma} & =0,\label{PSC-2}\\
(S_{2(\mu\nu)}+A_{2[\mu\nu]})+2(S'_{2(\mu\nu)}+A'_{2[\mu\nu]})+\sqrt{2\alpha'}
(2S_{3(\rho\mu\nu)}+B_{3(\rho[\mu)\nu]}+B_{3(\mu[\rho)\nu]})k^{\rho} & =0,\label{PSC-3}\\
\frac{3}{2}\zeta_{1\mu}+2\sqrt{2\alpha'}(S'_{2(\mu\nu)}-A'_{2[\mu\nu]})k^{\nu}
+(S_{3(\nu\rho\mu)}+B_{3(\nu[\rho)\mu]})\eta^{\nu\rho}&\nonumber \\ +~\frac{\eta^{\nu\rho}}{2}
(D'_{3[\nu(\mu]\rho)}-D'_{3[\mu(\nu]\rho)}) & =0.\label{PSC-4}
\end{align}
We are left with the third set of conditions. From $G_{\frac{5}{2}}|n=2\rangle=0$,
we obtain,
\begin{equation}
\sqrt{2\alpha'}\zeta_{1\mu}k^{\mu}+(S_{2(\mu\nu)}+2S'_{2(\mu\nu)})\eta^{\mu\nu}=0.\label{PSC-5}
\end{equation}
From $G_{\frac{3}{2}}|n=2\rangle=0$,
\begin{align}
\zeta_{1\mu}+\sqrt{2\alpha'}(S_{2(\mu\nu)}+A_{2[\mu\nu]})k^{\nu}+2S_{3(\mu\nu\rho)}\eta^{\nu\rho}+(B_{4(\mu[\nu)\rho]}+B_{4(\nu[\mu)\rho]})\eta^{\nu\rho} & =0,\label{PSC-6}\\
(B_{4(\rho[\sigma)\mu\nu]}-B_{4(\rho[\mu)\sigma\nu]}+B_{4(\rho[\mu)\nu\sigma]})\eta^{\rho\sigma}+2A'_{2[\mu\nu]}+\sqrt{2\alpha'}(D'_{3[\mu(\nu]\rho)}+A'_{3[\mu\nu\rho]})k^{\rho} & =0.\label{PSC-7}
\end{align}
Finally, $G_{\frac{1}{2}}|n=2\rangle=0$ yields,
\begin{align}
A_{4[\mu\nu\rho\sigma]} & =0,\label{PSC-8}\\
\zeta_{1\mu}+\sqrt{2\alpha'}(S'_{2(\mu\nu)}+A'_{2[\mu\nu]})k^{\nu} & =0,\label{PSC-9}\\
S_{2(\mu\nu)}+\sqrt{2\alpha'}(S_{3(\mu\nu\rho)}+B_{3(\mu[\nu)\rho]})k^{\rho} & =0,\label{PSC-10}\\
A_{2[\mu\nu]}+2A'_{2[\mu\nu]}+\sqrt{2\alpha'}(D'_{3[\rho(\mu]\nu)}-D'_{3[\mu(\rho]\nu)}+2A'_{3[\mu\nu\rho]})k^{\rho} & =0.\label{PSC-11}\\
3\sqrt{2\alpha'}B_{4(\mu[\nu)\rho\sigma]}k^{\sigma}+B_{3(\mu[\nu)\rho]}+\frac{1}{2}B_{3(\nu[\mu)\rho]}-\frac{1}{2}B_{3(\rho[\mu)\nu]}+D'_{3[\nu(\rho]\mu)}+A'_{3[\mu\nu\rho]} & =0,\label{PSC-12}
\end{align}

First, we take care of simplest conditions. We get $A_{4[\mu\nu\rho\sigma]}=0$
directly from Eq.(\ref{PSC-8}). Similarly, Eq.(\ref{PSC-12}) requires
$A'_{3[\mu\nu\rho]}=0$. Thus, Eq.(\ref{PSC-2}) now reads
\begin{equation}
B_{4(\sigma[\mu)\nu\rho]}k^{\sigma}=0.\label{B4Tran1}
\end{equation}
Furthermore, by examining all equations involving $B_{4(\mu[\nu\rho\sigma)}$,
we find the consistency condition $B_{4(\mu[\nu\rho\sigma)}k^{\sigma}=0$
which, together with Eq.(\ref{B4Tran1}), impose transversality of
$B_{4(\mu[\nu\rho\sigma)}$ with respect to all indices:
\begin{equation}
B_{4\mu_{1}\mu_{2}\mu_{3}\mu_{4}}k^{\mu_{i}}=0,\qquad(i=1,2,3,4).\label{B4Tran2}
\end{equation}
Notice that now, Eq.(\ref{PSC-12}) becomes
\begin{equation}
B_{3(\mu[\nu)\rho]}+\frac{1}{2}B_{3(\nu[\mu)\rho]}-\frac{1}{2}B_{3(\rho[\mu)\nu]}+D'_{3[\nu(\rho]\mu)}=0
\end{equation}

Next, Eq.(\ref{PSC-4}) splits into
\begin{align}
S_{2(\mu\nu)}+2S'_{2(\mu\nu)}+\sqrt{2\alpha'}2S_{3(\rho\mu\nu)}k^{\rho} & =0,\\
A_{2[\mu\nu]}+2A'_{2[\mu\nu]}+\sqrt{2\alpha'}(B_{3(\rho[\mu)\nu]}+B_{3(\mu[\rho)\nu]})k^{\rho} & =0.\label{ABrelation}
\end{align}
Note also that Eq.(\ref{PSC-11}) becomes
\begin{equation}
A_{2[\mu\nu]}+2A'_{2[\mu\nu]}+\sqrt{2\alpha'}(D'_{3[\rho(\mu]\nu)}-D'_{3[\mu(\rho]\nu)})k^{\rho}=0.\label{ADrelation}
\end{equation}
After multiplying both sides by $k^{\mu}$, we obtain
\begin{equation}
(A_{2[\mu\nu]}+2A'_{2[\mu\nu]})k^{\mu}=0.
\end{equation}

On the other hand, Eq.(\ref{PSC-4})$-\frac{1}{2}\times$Eq.(\ref{PSC-6})$-\frac{\eta^{\mu\nu}}{2}\times$Eq.(\ref{PSC-12})
gives us
\begin{equation}
\zeta_{1\mu}+2\sqrt{2\alpha'}(S'_{2(\mu\nu)}-A'_{2[\mu\nu]})k^{\nu}-\frac{1}{2}\sqrt{2\alpha'}(S_{2(\mu\nu)}+A_{2[\mu\nu]})k^{\nu}=0.
\end{equation}
After inserting this into Eq.(\ref{PSC-1}) and Eq.(\ref{PSC-9}) we find
\begin{equation}
-4A'_{2[\mu\nu]}k^{\nu}-A_{2[\mu\nu]}k^{\nu}=0.
\end{equation}
In this way, we obtain
\begin{equation}
A_{2[\mu\nu]}k^{\mu}=A'_{2[\mu\nu]}k^{\mu}=0.
\end{equation}

Taking into account all equations allows decoupling $\zeta_{1\mu}$,
$S_{2(\mu\nu)}$, $S'_{2(\mu\nu)}$ and $S_{3(\mu\nu\rho)}$ from
other fields. After removing all the dependent relations, we obtain
the following set:

\begin{equation}
\left\{ \begin{aligned} & 2\zeta_{1\mu}+\sqrt{2\alpha'}S_{2(\mu\nu)}k^{\nu}=0\\
 & S_{2(\mu\nu)}+\sqrt{2\alpha'}S_{3(\mu\nu\rho)}k^{\rho}=0\\
 & 2S'_{2(\mu\nu)}+\sqrt{2\alpha'}S_{3(\mu\nu\rho)}k^{\rho}=0\\
 & \sqrt{2\alpha'}S'_{2(\mu\nu)}k^{\nu}+2S_{3(\mu\nu\rho)}\eta^{\nu\rho}=0
\end{aligned}
\right..\label{2ndES-1}
\end{equation}
The solutions are enumerated below:
\begin{enumerate}
\item $S_{3(\mu\nu\rho)}=\sigma_{\mu\nu\rho}$ and $S_{2(\mu\nu)}=S'_{2(\mu\nu)}=\zeta_{1\mu}=0$.
$\sigma_{\mu\nu\rho}$ is a spin-3 field which satisfies
\begin{equation}
\sigma_{\mu\nu\rho}k^{\rho}=\sigma_{\mu\nu\rho}\eta^{\mu\nu}=0,
\end{equation}
and its vertex operator reads
\begin{equation}
V_{\sigma}=\frac{1}{2\alpha'}\sigma_{\mu\nu\rho}i\partial X^{\mu}i\partial X^{\nu}\psi^{\rho}e^{-\phi}e^{ikX}.
\end{equation}

\item $S_{3(\mu\nu\rho)}=\sqrt{2\alpha'}(\pi_{\mu\nu}k_{\rho}+\pi_{\mu\rho}k_{\nu}+\pi_{\nu\rho}k_{\mu})$,
$S_{2(\mu\nu)}=4\pi_{\mu\nu}$, $S'_{2(\mu\nu)}=2\pi_{\mu\nu}$, and
$\zeta_{1\mu}=0$, where $\pi_{\mu\nu}$ is a spin-2 field satisfying
$\pi_{\mu\nu}k^{\nu}=\pi_{\mu\nu}\eta^{\mu\nu}=0$. The corresponding spin-2
vertex operator is
\begin{align}
V_{\pi} & =\Big(\sqrt{\frac{1}{2\alpha'}}(\pi_{\mu\nu}k_{\rho}+\pi_{\mu\rho}
k_{\nu}+\pi_{\nu\rho}k_{\mu})i\partial X^{\mu}i\partial X^{\nu}\psi^{\rho}\nonumber \\
 & \quad+4\pi_{\mu\nu}\sqrt{\frac{1}{2\alpha'}}i\partial X^{\mu}\partial\psi^{\nu}+2\pi_{\mu\nu}\sqrt{\frac{1}{2\alpha'}}i\partial^{2}
 X^{\mu}\psi^{\nu}\Big)e^{-\phi}e^{ikX}.
\end{align}

\item $S_{3(\mu\nu\rho)}=\tilde{\zeta}_{3\mu\nu\rho}$, $S_{2(\mu\nu)}=\tilde{\zeta}_{2\mu\nu}$,
$S'_{2(\mu\nu)}=\tilde{\zeta}'_{2\mu\nu}$ and $\zeta_{1\mu}=\tilde{\zeta}_{1\mu}$,
where
\begin{align}
\tilde{\zeta}_{3\mu\nu\rho} & =\eta_{\mu\nu}\xi_{\rho}+\eta_{\mu\rho}\xi_{\nu}+\eta_{\nu\rho}\xi_{\mu}+
c(2\alpha')(k_{\mu}k_{\nu}\xi_{\rho}+k_{\mu}\xi_{\nu}k_{\rho}+\xi_{\mu}k_{\nu}k_{\rho}),\\
\tilde{\zeta}_{2\mu\nu} & =(4c-1)\sqrt{2\alpha'}(k_{\mu}\xi_{\nu}+\xi_{\mu}k_{\nu}),\\
\tilde{\zeta}'_{2\mu\nu} & =\frac{1}{2}(4c-1)\sqrt{2\alpha'}(k_{\mu}\xi_{\nu}+\xi_{\mu}k_{\nu}),\\
\tilde{\zeta}_{1\mu} & =2(4c-1)\xi_{\mu},
\end{align}
with $c=7/8$. $\xi_{\mu}$ is a vector field satisfying $\xi_{\mu}k^{\mu}=0$.
The corresponding vector vertex operator reads
\begin{align}
V_{\xi}^{(1)} & =\Big(\tilde{\zeta}_{3\mu\nu\rho}\frac{1}{2\alpha'}i\partial X^{\mu}i\partial X^{\nu}\psi^{\rho}+\tilde{\zeta}_{2\mu\nu}\sqrt{\frac{1}{2\alpha'}}i\partial X^{\mu}\partial\psi^{\nu}\nonumber \\
 & \quad+\tilde{\zeta}'_{2\mu\nu}\sqrt{\frac{1}{2\alpha'}}i\partial^{2}
 X^{\mu}\psi^{\nu}+\tilde{\zeta}_{1\mu}\frac{1}{2}\partial^{2}\psi^{\mu}
 \Big)e^{-\phi}e^{ikX}.\label{pv-1}
\end{align}

\item $S_{3\mu\nu\rho}=\hat{\zeta}_{3\mu\nu\rho}$, $S_{2\mu\nu}=\hat{\zeta}_{2\mu\nu}$,
$S'_{2\mu\nu}=\hat{\zeta}'_{2\mu\nu}$ and $\zeta_{1\mu}=\hat{\zeta}_{1\mu}$,
where
\begin{flalign}
\hat{\zeta}_{3\mu\nu\rho} & =\big[\sqrt{2\alpha'}(\eta_{\mu\nu}k_{\rho}+\eta_{\mu\rho}k_{\nu}+
\eta_{\nu\rho}k_{\mu})+d(2\alpha')^{\frac{3}{2}}k_{\mu}k_{\nu}k_{\rho}\big]\varphi,\\
\hat{\zeta}_{2\mu\nu} & =\big[4\eta_{\mu\nu}-2\alpha'(2-4d)k_{\mu}k_{\nu}\big]\varphi,\\
\hat{\zeta}'_{2\mu\nu} & =\big[2\eta_{\mu\nu}-\alpha'(2-4d)k_{\mu}k_{\nu}\big]\varphi,\\
\hat{\zeta}_{1\mu} & =\big[-2(3-4d)\sqrt{2\alpha'}k_{\mu}\big]\varphi,
\end{flalign}
with $d=9/8$. $\varphi$ is a scalar field, and its vertex operator
is
\begin{align}
V_{\varphi} & =\Big(\hat{\zeta}_{3\mu\nu\rho}\frac{1}{2\alpha'}i\partial X^{\mu}i\partial X^{\nu}\psi^{\rho}+\hat{\zeta}_{2\mu\nu}\sqrt{\frac{1}{2\alpha'}}i\partial X^{\mu}\partial\psi^{\nu}\nonumber \\
 & \quad+\hat{\zeta}'_{2\mu\nu}\sqrt{\frac{1}{2\alpha'}}i\partial^{2}
 X^{\mu}\psi^{\nu}+\hat{\zeta}_{1\mu}\frac{1}{2}\partial^{2}\psi^{\mu}\Big)e^{-\phi}e^{ikX}.
\end{align}

\end{enumerate}
Let us check if we identified all independents degrees of freedom. We started from a totally symmetric
3-tensor $S_3$ (20 d.o.f.), two symmetric 2-tensors $S_2$ and $S_2'$ (20 d.o.f.) and one
vector $\xi_1$ (4 d.o.f.), a total  of 44 d.o.f. The set (\ref{2ndES-1}) contains
$4+10+10+4=28$ constraints. Thus we are left with 16 independent d.o.f., which
are one spin-3 field $\sigma$ (7 d.o.f.), one spin-2 field $\pi$ (5 d.o.f.),
one vector $\xi$ (3 d.o.f.) and one scalar $\varphi$ (1 d.o.f.).
Next, we examine the vertex operators to check if any of the above degrees of freedom happens to represent a null state. Indeed, we find that the spin-2 field $\pi$ and the scalar $\varphi$ are null states.
They do not couple to two massless gluons in any helicity configurations,
just like the two vectors found at the first massive level, {\em c.f}.\ Eqs.(\ref{1stNv1})
and (\ref{1stNv2}).

Now we turn to remaining fields. With our previous analysis, after eliminating all the dependent relations, we arrive to
the following set:
\begin{equation}
\left\{ \begin{aligned} & (B_{4(\rho[\sigma)\mu\nu]}-B_{4(\rho[\mu)\sigma\nu]}+B_{4(\rho[\mu)\nu\sigma]})\eta^{\rho\sigma}+2A'_{2[\mu\nu]}+\sqrt{2\alpha'}D'_{3[\mu(\nu]\rho)}k^{\rho}=0\\
 & B_{4(\mu[\nu)\rho\sigma]}k^{\mu}=B_{4(\mu[\nu)\rho\sigma]}k^{\sigma}=0\\
 & A_{2[\mu\nu]}+2A'_{2[\mu\nu]}+\sqrt{2\alpha'}(B_{3(\rho[\mu)\nu]}+B_{3(\mu[\rho)\nu]})k^{\rho}=0\\
 & A_{2[\mu\nu]}+2A'_{2[\mu\nu]}+\sqrt{2\alpha'}(D'_{3[\rho(\mu]\nu)}-D'_{3[\mu(\rho]\nu)})k^{\rho}=0\\
 & B_{3(\mu[\nu)\rho]}+\frac{1}{2}B_{3(\nu[\mu)\rho]}-\frac{1}{2}B_{3(\rho[\mu)\nu]}+D'_{3[\nu(\rho]\mu)}=0\\
 & B_{3(\mu[\nu)\rho]}k^{\rho}=0
\end{aligned}
\right..\label{2ndES-2}
\end{equation}
The solutions  are:
\begin{enumerate}
\item $B_{3\mu\nu\rho}=\eta_{\mu\nu}^{\bot}\xi_{\rho}-\frac{1}{4}\xi_{\mu}\eta_{\nu\rho}^{\bot}-\frac{1}{4}\xi_{\nu}\eta_{\mu\rho}^{\bot}$,
$D'_{3\mu\nu\rho}=\frac{1}{2}\xi_{\mu}\eta_{\nu\rho}^{\bot}-2\xi_{\nu}\eta_{\mu\rho}^{\bot}$,
$B_{4}=A_{2}=A'_{2}=0$, where $\xi_{\mu}$ is a spin-1 wavefunction satisfying
$\xi_{\mu}k^{\mu}=0$.
\item $B_{3\mu\nu\rho}=-\frac{1}{2}D'_{3\mu\nu\rho}=k^{\sigma}\varepsilon_{\sigma\mu\rho\gamma}\pi'^{\gamma\lambda}\eta_{\lambda\nu}+k^{\sigma}\varepsilon_{\sigma\nu\rho\gamma}\pi'^{\gamma\lambda}\eta_{\lambda\mu}$,
$B_{4}=A_{2}=A'_{2}=0$, where $\pi'_{\mu\nu}$ is another spin-2
field satisfying $\pi'_{\mu\nu}k^{\nu}=\pi'_{\mu\nu}\eta^{\mu\nu}=0$.
\item $B_{4\mu\nu\rho\sigma}=x\upsilon_{(\mu}E_{\nu)\rho\sigma},$
\footnote{The choice of the wave function of $B_{4\mu\nu\rho\sigma}$ is not
unique. Indeed we find another solution with $B_{4\mu\nu\rho\sigma}=2\eta_{\mu\nu}^{\bot}\upsilon^{\tau}E_{\tau\rho\sigma}$,
where $\eta_{\mu\nu}^{\bot}\equiv\eta_{\mu\nu}-k_{\mu}k_{\nu}/k^{2}$
and $E_{\mu\nu\rho}$ is the same 3-form. However, this solution
does not give us an extra physical field. When we compute the scattering
amplitudes involving the physical fields subject to these two solutions,
we find the results are exactly the same. That is to say, these two
different solutions represent the same physical vector state. We will
stick to the first solution in our later discussions.%
} $B_{3\mu\nu\rho}=y\sqrt{2\alpha'}k_{\mu}\upsilon^{\tau}E_{\tau\nu\rho}$,
$D'_{3\mu\nu\rho}=-y\sqrt{2\alpha'}(\upsilon^{\tau}E_{\tau\mu\nu}k_{\rho}+\frac{1}{2}\upsilon^{\tau}E_{\tau\rho\nu}k_{\mu}-\frac{1}{2}\upsilon^{\tau}E_{\tau\rho\mu}k_{\nu})$,
$A'_{2\mu\nu}=-(x+2y)\upsilon^{\tau}E_{\tau\mu\nu}$, $A_{2\mu\nu}=(2x+8y)\upsilon^{\tau}E_{\tau\mu\nu}$,
where the vector $\upsilon$ is transverse, $\upsilon_{\mu}k^{\mu}=0$, and
the 3-form $E_{\mu\nu\rho}=\frac{i}{6}\sqrt{2\alpha'}\varepsilon_{\mu\nu\rho\sigma}k^{\sigma}$.
Although only one massive vector field $\upsilon$ is involved in
our solution, we still have two parameters $x$, $y$ available, thus
we get two pseudo-vectors, $\upsilon_{\mu}(x_{1},y_{1})$ and $\upsilon_{\mu}(x_{2},y_{2})$.
There is a natural choice for the coefficients $x$, $y$, dictated
by the complexification of vector fields, to be made after discussing
the helicity-dependence of their couplings to gauge bosons.
\end{enumerate}
Let us count the number of degrees of freedom again. We started from
a hook 4-index tensor $B_{4}$ with 15 d.o.f., two hook 3-tensors,
$B_{3}$ and $D'_{3}$, with $20\times2=40$ d.o.f., and two antisymmetric
2-tensors $A_{2}$, $A'_{2}$ with $6\times2=12$ d.o.f. The set (\ref{2ndES-2})
contains $6+12+6+6+20+6=56$ constraints%
\footnote{Relation $B_{4(\mu[\nu)\rho\sigma]}k^{\mu}=B_{4(\mu[\nu)\rho\sigma]}k^{\sigma}=0$
give total $8+4=12$ constraints. First of all $B_{4(\mu[\nu)\rho\sigma]}k^{\mu}=0$
kills the second box in the first row, so we are left with a Young
diagram $\tiny{\yng(1,1,1)}$, which corresponds to $\frac{4\times3\times2}{3\times2}=4$
constraints. In addition, $B_{4(\mu[\nu)\rho\sigma]}k^{\sigma}=0$
kills the box in the third row, so we are left with $\tiny{\yng(2,1)}$,
which gives us $\frac{3\times4\times2}{3}=8$ more constraints. The
subtlety here is when one of the antisymmetric indices is eliminated,
once we calculate the dimensions of the Young diagram, the number
we put in the first box is 3 instead of 4.%
}$^{,}$%
\footnote{Similarly to the previous case, $B_{3(\mu[\nu)\rho]}k^{\rho}=0$ represents
6 constraints. $B_{3(\mu[\nu)\rho]}k^{\rho}=0$ corresponds to $\tiny{\yng(2)}$
so it gives us $\frac{3\times4}{2}=6$ constraints. Note that again
the number we put in the first box is 3 instead of 4 because one of
the antisymmetric indices is eliminated.%
}. We are left with 11 d.o.f.: two vector fields $\xi_{\mu}$ and $\upsilon_{\mu}$
(3 d.o.f. each) and one spin-2 field $\pi'$ (5 d.o.f.).

The vertex operators of these physical fields are:
\begin{align}
V_{\xi}^{(2)} & =\Big[\xi_{\mu}\eta_{\nu\rho}^{\bot}\frac{1}{2\alpha'}(\psi^{\mu}i\partial X^{\nu}i\partial X^{\rho}-\frac{1}{2}i\partial X^{\mu}i\partial X^{\nu}\psi^{\rho})+\frac{5}{2}\xi_{\mu}\eta_{\nu\rho}^{\bot}\psi^{\mu}\psi^{\nu}\partial\psi^{\rho}\Big]e^{-\phi}e^{ikX},\label{PTV-2}\\
V_{\pi'} & =(k^{\sigma}\varepsilon_{\sigma\mu\rho\gamma}\pi'^{\gamma\lambda}\eta_{\lambda\nu}+k^{\sigma}\varepsilon_{\sigma\nu\rho\gamma}\pi'^{\gamma\lambda}\eta_{\lambda\mu})\Big[\big(\frac{1}{2\alpha'}\big)i\partial X^{\mu}i\partial X^{\nu}\psi^{\rho}-2\partial\psi^{\mu}\psi^{\nu}\psi^{\rho}\Big]e^{-\phi}e^{ikX},
\end{align}
and
\begin{align}
V_{\upsilon(x,y)} & =\frac{x}{\sqrt{2\alpha'}}\Big[\big(\upsilon^{\tau}E_{\tau\mu\nu}i\partial^{2}X^{\mu}\psi^{\nu}-2\upsilon^{\tau}E_{\tau\mu\nu}i\partial X^{\mu}\partial\psi^{\nu}\big)+\upsilon_{(\mu}E_{\nu)\rho\sigma}i\partial X^{\mu}\psi^{\nu}\psi^{\rho}\psi^{\sigma}\Big]e^{-\phi}e^{ikX}\nonumber \\
 & \,+\frac{y}{\sqrt{2\alpha'}}\Big[k_{\mu}\upsilon^{\tau}E_{\tau\nu\rho}i\partial X^{\mu}i\partial X^{\nu}\psi^{\rho}-2(2\alpha')\upsilon^{\tau}E_{\tau\mu(\nu}k_{\rho)}\psi^{\mu}\psi^{\nu}\partial\psi^{\rho}\nonumber \\
 & \qquad\qquad\;+\big(8\upsilon^{\tau}E_{\tau\mu\nu}i\partial X^{\mu}\partial\psi^{\nu}-2\upsilon^{\tau}E_{\tau\mu\nu}i\partial^{2}X^{\mu}\psi^{\nu}\big)\Big]e^{-\phi}e^{ikX}.\label{PSV}
\end{align}

To summarize, we identified one spin-3 field, two spin-2 fields, four
vector fields (two vectors and two pseudo-vectors), and one real scalar
satisfying the physical state conditions. The scalar $\varphi$ and
the spin-2 field $\phi$ are null states; all other fields are physical.
Thus the number of universal physical degrees of freedom at the second level
of NS sector is 24. The spin-3 field and the spin-2 field $\pi$ couple
to two massless gluons with opposite helicities -- these are the particles responsible for
 the Regge pole in Eq.(\ref{2nd+-}). The four spin-1 fields
will pair up to form two complex vectors that can decay into
two gluons with the same helicities only, {\em c.f}.\ the pole in Eq.(\ref{2ndVector}).

Both even- and odd-parity particles couple to $(++)$ and $(--)$ gluon helicity configurations. The relative normalization of their couplings is dictated by supersymmetry which forbids non-vanishing ``all-plus'' and ``all-minus'' scattering amplitudes \cite{SUSYRe}. Thus similarly to the scalar $\Phi$ at the first level, the vectors and pseudo-vectors  of the second level must combine to form complex vector fields that couple to gluons with the selection rules similar to Eq.(\ref{1stScalarSR}). To that end, we introduce two complex vector fields, $\Xi_{1,2}^{\pm}$, with the vertices
\begin{align}
V_{\Xi_{1}^{\pm}} & =V_{\xi}^{(1)}\pm V_{\upsilon(x_{1},y_{1})}(\xi)\nonumber \\
 & =CT^{a}\Big\{\big[\big(\frac{3}{2\alpha'}\xi_{(\mu}\eta_{\nu\rho)}+\frac{21}{8}\xi_{(\mu}k_{\nu}k_{\rho)}\big)i\partial X^{\mu}i\partial X^{\nu}\psi^{\rho}\nonumber \\
 & \qquad\qquad+5\xi_{(\mu}k_{\nu)}i\partial X^{\mu}\partial\psi^{\nu}+\frac{5}{2}\xi_{(\mu}k_{\nu)}i\partial^{2}X^{\mu}\psi^{\nu}+\frac{5}{2}\xi_{\mu}\partial^{2}\psi^{\mu}\big]\nonumber \\
 & \quad\pm\big\{\frac{x_{1}}{\sqrt{2\alpha'}}\big[\big(\xi^{\tau}E_{\tau\mu\nu}i\partial^{2}X^{\mu}\psi^{\nu}-2\xi^{\tau}E_{\tau\mu\nu}i\partial X^{\mu}\partial\psi^{\nu}\big)+\xi_{(\mu}E_{\nu)\rho\sigma}i\partial X^{\mu}\psi^{\nu}\psi^{\rho}\psi^{\sigma}\big]\nonumber \\
 & \quad\;\;+\frac{y_{1}}{\sqrt{2\alpha'}}\big[k_{\mu}\xi^{\tau}E_{\tau\nu\rho}i\partial X^{\mu}i\partial X^{\nu}\psi^{\rho}-2(2\alpha')\xi^{\tau}E_{\tau\mu(\nu}k_{\rho)}\psi^{\mu}\psi^{\nu}\partial\psi^{\rho}\nonumber \\
 & \qquad\qquad\quad\;+\big(8\xi^{\tau}E_{\tau\mu\nu}i\partial X^{\mu}\partial\psi^{\nu}-2\xi^{\tau}E_{\tau\mu\nu}i\partial^{2}X^{\mu}\psi^{\nu}\big)\big]\big\}\Big\} e^{-\phi}e^{ikX}.\label{firstv}
\end{align}
\begin{align}
V_{\Xi_{2}^{\pm}} & =V_{\xi}^{(2)}\pm V_{\upsilon(x_{2},y_{2})}(\xi)\nonumber \\
 & =CT^{a}\Big\{\big[\xi_{\mu}\eta_{\nu\rho}^{\bot}\frac{1}{2\alpha'}(\psi^{\mu}i\partial X^{\nu}i\partial X^{\rho}-\frac{1}{2}i\partial X^{\mu}i\partial X^{\nu}\psi^{\rho})+\frac{5}{2}\xi_{\mu}\eta_{\nu\rho}^{\bot}\psi^{\mu}\psi^{\nu}
 \partial\psi^{\rho}\big]\nonumber \\
 & \quad\pm\big\{\frac{x_{2}}{\sqrt{2\alpha'}}\big[\big(\xi^{\tau}E_{\tau\mu\nu}i\partial^{2}X^{\mu}\psi^{\nu}-2\xi^{\tau}E_{\tau\mu\nu}i\partial X^{\mu}\partial\psi^{\nu}\big)+\xi_{(\mu}E_{\nu)\rho\sigma}i\partial X^{\mu}\psi^{\nu}\psi^{\rho}\psi^{\sigma}\big]\nonumber \\
 & \quad\;\;+\frac{y_{2}}{\sqrt{2\alpha'}}\big[k_{\mu}\xi^{\tau}E_{\tau\nu\rho}i\partial X^{\mu}i\partial X^{\nu}\psi^{\rho}-2(2\alpha')\xi^{\tau}E_{\tau\mu(\nu}k_{\rho)}\psi^{\mu}\psi^{\nu}\partial\psi^{\rho}\nonumber \\
 & \qquad\qquad\quad\;+\big(8\xi^{\tau}E_{\tau\mu\nu}i\partial X^{\mu}\partial\psi^{\nu}-2\xi^{\tau}E_{\tau\mu\nu}i\partial^{2}X^{\mu}\psi^{\nu}\big)\big]\big\}\Big\} e^{-\phi}e^{ikX}.\label{secondv}
\end{align}
The coefficients $(x_{1},y_{1})$ and $(x_{2},y_{2})$ will be fixed by requiring that
$\Xi_{1,2}^{+}$ couple to two gluons in (++) configurations and to three gluons in mostly plus configurations only (at least two gluons carrying positive helicities). The overall normalization factors $C$ will be fixed by the usual factorization arguments.

\subsection{Complex Vector Couplings to Two Gluons}

Three-point amplitudes with one massive vector (with the momentum and color indices labeled by 1) and two massless gluons are very simple because the positions of three vertices can be fixed by
using $PSL(2,R)$ invariance of the disk world-sheet and there are no integrals involved in the computations.
The three-point amplitude of the pseudo-vector (vertex $V_{\upsilon(x,y)}$)
and two  gluons reads
\begin{align}
\mathscr{A}^{(3)}(\upsilon_{(x,y)}(\xi),\epsilon_{2},\epsilon_{3})
 & ~=~C_{D_{2}}C\sqrt{2\alpha'}g^{2}f^{a_{1}a_{2}a_{3}}(2\alpha')^{\frac{3}{2}}\Big\{4x\varepsilon_{\mu\nu\rho}\epsilon_{2}^{\mu}\epsilon_{3}^{\nu}k_{3}^{\rho}(\xi\cdot k_{3})\nonumber \\
 & -(x+2y)\xi^{\rho}\varepsilon_{\rho\mu\nu}\big[\epsilon_{2}^{\mu}k_{3}^{\nu}(\epsilon_{3}\cdot k_{2})+\epsilon_{3}^{\mu}k_{3}^{\nu}(\epsilon_{2}\cdot k_{3})+\frac{1}{\alpha'}\epsilon_{2}^{\mu}\epsilon_{3}^{\nu}\big]\Big\}.
\end{align}
where $C_{D_2}=g^{-2}\alpha'^{-2}$ is the universal disk factor \cite{StringPol}.
In the helicity basis, this corresponds to
\begin{align}
\mathscr{A}^{(3)}(\xi,+,+) & =(\frac{x}{6}-\frac{y}{3})C_{D_{2}}C\sqrt{2\alpha'}g^{2}f^{a_{1}a_{2}a_{3}}(2\alpha')^{2}[23]^{2}(\xi\cdot k_{2}),\\
\mathscr{A}^{(3)}(\xi,+,-) & =0,\\
\mathscr{A}^{(3)}(\xi,-,-) & =-(\frac{x}{6}-\frac{y}{3})C_{D_{2}}C\sqrt{2\alpha'}g^{2}f^{a_{1}a_{2}a_{3}}(2\alpha')^{2}\langle23\rangle^{2}(\xi\cdot k_{2}).
\end{align}
The three-point amplitude of the vector $\xi_{(1)}$
(vertex $V_{\xi}^{(1)}$) and two  gluons is
\begin{align}
\mathscr{A}^{(3)}(\xi_{(1)},\epsilon_{2},\epsilon_{3})
 & =C_{D_{2}}C\sqrt{2\alpha'}g^{2}f^{a_{1}a_{2}a_{3}}(2\alpha')^{2}\Big\{\big(3\xi_{(\mu}\eta_{\nu\rho)}+\frac{21}{4}\alpha'\xi_{(\mu}k_{\nu}k_{\rho)}\big)\big[\frac{1}{2\alpha'}\epsilon_{2}^{\mu}\epsilon_{3}^{\nu}k_{3}^{\rho}\nonumber \\
 & -\frac{1}{2\alpha'}\epsilon_{2}^{\mu}\epsilon_{3}^{\nu}k_{2}^{\rho}-\epsilon_{2}^{\mu}k_{2}^{\nu}k_{3}^{\rho}(\epsilon_{3}\cdot k_{2})-\epsilon_{3}^{\mu}k_{2}^{\nu}k_{2}^{\rho}(\epsilon_{2}\cdot k_{3})+k_{2}^{\mu}k_{2}^{\nu}k_{3}^{\rho}(\epsilon_{2}\cdot\epsilon_{3})\big]\nonumber\\
 & +\frac{5}{2}\xi_{(\mu}k_{\nu)}\big[\epsilon_{2}^{\mu}k_{3}^{\nu}(\epsilon_{3}\cdot k_{2})+\epsilon_{3}^{\mu}k_{2}^{\nu}(\epsilon_{2}\cdot k_{3})-k_{2}^{\mu}k_{3}^{\nu}(\epsilon_{2}\cdot\epsilon_{3})\big]\Big\}.
\end{align}
The corresponding helicity amplitudes are
\begin{align}
\mathscr{A}^{(3)}(\xi_{(1)},+,+) & =\frac{5}{8}C_{D_{2}}C\sqrt{2\alpha'}g^{2}f^{a_{1}a_{2}a_{3}}(2\alpha')^{2}[23]^{2}(\xi\cdot k_{2}),\\
\mathscr{A}^{(3)}(\xi_{(1)},+,-) & =0,\\
\mathscr{A}^{(3)}(\xi_{(1)},-,-) & =\frac{5}{8}C_{D_{2}}C\sqrt{2\alpha'}g^{2}f^{a_{1}a_{2}a_{3}}(2\alpha')^{2}\langle23\rangle^{2}(\xi\cdot k_{2}).
\end{align}
The three-point amplitude of the  vector $\xi_{(2)}$
(vertex $V_{\xi}^{(2)}$) and two  gluons is
\begin{align}
\mathscr{A}^{(3)}(\xi_{(2)},\epsilon_{2},\epsilon_{3})
 & =C_{D_{2}}C\sqrt{2\alpha'}g^{2}f^{a_{1}a_{2}a_{3}}(2\alpha')^{2}\Big\{\big[(\xi\cdot\epsilon_{2})((\eta_{\mu\nu}^{\bot}k_{3}^{\mu}k_{3}^{\nu})(\epsilon_{3}\cdot k_{2})+(\xi\cdot k_{3})((\eta_{\mu\nu}^{\bot}k_{3}^{\mu}k_{3}^{\nu})(\epsilon_{2}\cdot\epsilon_{3})\nonumber \\
 & -(\xi\cdot\epsilon_{3})((\eta_{\mu\nu}^{\bot}k_{3}^{\mu}k_{3}^{\nu})(\epsilon_{2}\cdot k_{3})+\frac{1}{\alpha'}(\xi\cdot\epsilon_{2})((\eta_{\mu\nu}^{\bot}\epsilon_{3}^{\mu}k_{3}^{\nu})\big]-\frac{1}{2}\big[(\xi\cdot k_{3})(\eta_{\mu\nu}^{\bot}\epsilon_{2}^{\mu}k_{3}^{\nu})(\epsilon_{3}\cdot k_{2})\nonumber \\
 & +(\xi\cdot k_{3})((\eta_{\mu\nu}^{\bot}k_{3}^{\mu}k_{3}^{\nu})(\epsilon_{2}\cdot\epsilon_{3})-(\xi\cdot k_{3})((\eta_{\mu\nu}^{\bot}\epsilon_{3}^{\mu}k_{3}^{\nu})(\epsilon_{2}\cdot k_{3})+\frac{1}{2\alpha'}(\xi\cdot k_{3})(\eta_{\mu\nu}^{\bot}\epsilon_{2}^{\mu}\epsilon_{3}^{\nu})\nonumber \\
 & +\frac{1}{2\alpha'}(\xi\cdot\epsilon_{3})(\eta_{\mu\nu}^{\bot}\epsilon_{2}^{\mu}k_{3}^{\nu})\big]+\frac{5}{2}\frac{1}{2\alpha'}\big[(\xi\cdot\epsilon_{3})(\eta_{\mu\nu}^{\bot}\epsilon_{2}^{\mu}k_{3}^{\nu})-(\xi\cdot k_{3})(\eta_{\mu\nu}^{\bot}\epsilon_{2}^{\mu}\epsilon_{3}^{\nu})\big]\Big\}.
\end{align}
The corresponding helicity amplitudes are
\begin{align}
\mathscr{A}^{(3)}(\xi_{(2)},+,+) & =-\frac{5}{8}C_{D_{2}}C\sqrt{2\alpha'}g^{2}f^{a_{1}a_{2}a_{3}}(2\alpha')^{2}[23]^{2}(\xi\cdot k_{2}),\\
\mathscr{A}^{(3)}(\xi_{(2)},+,-) & =0,\\
\mathscr{A}^{(3)}(\xi_{(2)},-,-) & =-\frac{5}{8}C_{D_{2}}C\sqrt{2\alpha'}g^{2}f^{a_{1}a_{2}a_{3}}(2\alpha')^{2}\langle23\rangle^{2}(\xi\cdot k_{2}).
\end{align}

In the basis of complex vectors $\Xi_{1,2}^{\pm}(x,y)$, the above amplitudes correspond to
\begin{align}
\mathscr{A}^{(3)}(\Xi_{1}^{\pm}(\xi),+,+) & =\big[\frac{5}{8}\pm(\frac{x_{1}}{6}-\frac{y_{1}}{3})\big]C_{D_{2}}C\sqrt{2\alpha'}g^{2}f^{a_{1}a_{2}a_{3}}(2\alpha')^{2}[23]^{2}(\xi\cdot k_{2}),\label{3p++}\\
\mathscr{A}^{(3)}(\Xi_{1}^{\pm}(\xi),+,-) & =0,\\
\mathscr{A}^{(3)}(\Xi_{1}^{\pm}(\xi),-,-) & =\big[\frac{5}{8}\mp(\frac{x_{1}}{6}-\frac{y_{1}}{3})\big]C_{D_{2}}C\sqrt{2\alpha'}g^{2}f^{a_{1}a_{2}a_{3}}(2\alpha')^{2}\langle23\rangle^{2}(\xi\cdot k_{2}).\label{3p--}
\end{align}
and
\begin{align}
\mathscr{A}^{(3)}(\Xi_{2}^{\pm}(\xi),+,+) & =\big[-\frac{5}{8}\pm(\frac{x_{2}}{6}-\frac{y_{2}}{3})\big]C_{D_{2}}C\sqrt{2\alpha'}g^{2}f^{a_{1}a_{2}a_{3}}(2\alpha')^{2}[23]^{2}(\xi\cdot k_{2}),\\
\mathscr{A}^{(3)}(\Xi_{2}^{\pm}(\xi),+,-) & =0,\\
\mathscr{A}^{(3)}(\Xi_{2}^{\pm}(\xi),-,-) & =\big[-\frac{5}{8}\mp(\frac{x_{2}}{6}-\frac{y_{2}}{3})\big]C_{D_{2}}C\sqrt{2\alpha'}g^{2}f^{a_{1}a_{2}a_{3}}(2\alpha')^{2}\langle23\rangle^{2}(\xi\cdot k_{2}).
\end{align}
By requiring that $\Xi_{1,2}^{+}$
couple to $(+,+)$ only (and respectively, $\Xi_{1,2}^{-}$
to $(-,-)$ only), we obtain the following constraints:
\begin{align}
\frac{5}{8}-(\frac{x_{1}}{6}-\frac{y_{1}}{3}) & =0,\label{const1}\\
-\frac{5}{8}-(\frac{x_{2}}{6}-\frac{y_{2}}{3}) & =0.\label{const2}
\end{align}

\subsection{Complex Vector Couplings to Three Gluons}

We consider four-point amplitudes involving one massive vector and three gluons.\footnote{The original four-point string amplitudes are very tedious, so we will only present the helicity amplitudes in this paper, which look much simpler.} The kinematic variables are defined in Eq.(\ref{mandel}). Now, $k_1$ is the momentum of the massive particle, $k_1^2=2M^2$, and the Mandelstam variables satisfy
\begin{equation}
s+t+u=2M^2=\frac{2}{\alpha'}.
\end{equation}
All other quantum numbers associated to the massive vector will be also labeled by 1.

We begin with the amplitudes involving three all-plus and all-minus gluons,$(+++)$
and $(---)$, respectively. For all-plus configurations, they contain the common factors
\begin{align}
\mathscr{F}(j_z=+1,+,+,+) & =C_{D_{2}}C\sqrt{2\alpha'}g^{3}\mathscr{T}_{n=2}^{a_{1}a_{2}a_{3}a_{4}}V_{t}(2\alpha')^{3}\frac{\langle qp\rangle}{m}\Big\{\frac{(1-\alpha'u)}{\alpha'^{2}}\frac{[2q][q3]}{\langle34\rangle\langle42\rangle}\nonumber \\
 & \qquad+\frac{(1-\alpha's)}{\alpha'^{2}}\frac{[3q][q4]}{\langle23\rangle\langle42\rangle}+\frac{(1-\alpha't)}{\alpha'^{2}}\frac{[4q][q2]}{\langle23\rangle\langle34\rangle}\Big\},
\end{align}
\begin{align}
\mathscr{F}(j_z=0,+,+,+) & =C_{D_{2}}C\sqrt{2\alpha'}g^{3}\mathscr{T}_{n=2}^{a_{1}a_{2}a_{3}a_{4}}V_{t}(2\alpha')^{3}\frac{\langle qp\rangle}{\sqrt{2}m}\Big\{\frac{(1-\alpha'u)}{\alpha'^{2}}\frac{\big([2q][p3]+[3q][p2]\big)}{\langle34\rangle\langle42\rangle}\nonumber \\
 & +\frac{(1-\alpha's)}{\alpha'^{2}}\frac{\big([3q][p4]+[4q][p3]\big)}{\langle23\rangle\langle42\rangle}+\frac{(1-\alpha't)}{\alpha'^{2}}\frac{\big([4q][p2]+[2q][p4]\big)}{\langle23\rangle\langle34\rangle}\Big\},
\end{align}
\begin{align}\mathscr{F}(j_z=-1,+,+,+)
 & =C_{D_{2}}C\sqrt{2\alpha'}g^{3}\mathscr{T}_{n=2}^{a_{1}a_{2}a_{3}a_{4}}V_{t}(2\alpha')^{3}\frac{\langle qp\rangle}{m}\Big\{\frac{(1-\alpha'u)}{\alpha'^{2}}\frac{[2p][p3]}{\langle34\rangle\langle42\rangle}\nonumber \\
 & \qquad+\frac{(1-\alpha's)}{\alpha'^{2}}\frac{[3p][p4]}{\langle23\rangle\langle42\rangle}+\frac{(1-\alpha't)}{\alpha'^{2}}\frac{[4p][p2]}{\langle23\rangle\langle34\rangle}\Big\}.
\end{align}
Here, $\mathscr{T}_{n=2}^{a_{1}a_{2}a_{3}a_{4}}$ is a universal factor that combines Chan-Paton factors with the kinematic variables in the following way
\begin{align}
\mathscr{T}_{n=2}^{a_{1}a_{2}a_{3}a_{4}} & ={\rm Tr}(T^{a_{1}}T^{a_{2}}T^{a_{3}}T^{a_{4}}+T^{a_{4}}T^{a_{3}}T^{a_{2}}T^{a_{1}})\nonumber \\
 & +\frac{1}{V_{t}}\frac{V_{s}}{\alpha's-1}{\rm Tr}(T^{a_{2}}T^{a_{3}}T^{a_{1}}T^{a_{4}}+T^{a_{4}}T^{a_{1}}T^{a_{3}}T^{a_{2}})\nonumber \\
 & +\frac{1}{V_{t}}\frac{V_{u}}{\alpha'u-1}{\rm Tr}(T^{a_{3}}T^{a_{1}}T^{a_{2}}T^{a_{4}}+T^{a_{4}}T^{a_{2}}T^{a_{1}}T^{a_{3}}).
\end{align}
Furthermore, $p$ and $q$ are the light-like reference vectors used to define the quantization axis for the polarization vector $\xi$, see Appendix C.
We obtain
\begin{align}
\mathscr{A}^{(4)}(\Xi_{1}^{\pm}(j_z),+,+,+) & =\big[\frac{5}{8}\pm(\frac{x_{1}}{6}-\frac{y_{1}}{3})\big]
\mathscr{F}(j_z,+,+,+),\\
\mathscr{A}^{(4)}(\Xi_{2}^{\pm}(j_z),+,+,+) & =\big[-\frac{5}{8}\pm(\frac{x_{2}}{6}-\frac{y_{2}}{3})\big]
\mathscr{F}(j_z,+,+,+)
\end{align}
{}For all-minus configurations, the analogous expressions read
\begin{align}
\mathscr{F}(j_z=+1,-,-,-) & =C_{D_{2}}C\sqrt{2\alpha'}g^{3}\mathscr{T}_{n=2}^{a_{1}a_{2}a_{3}a_{4}}V_{t}(2\alpha')^{3}\frac{[pq]}{m}\times\Big\{\frac{(1-\alpha'u)}{\alpha'^{2}}\frac{\langle2p\rangle\langle p3\rangle}{[34][42]}\nonumber \\
 & \qquad+\frac{(1-\alpha's)}{\alpha'^{2}}\frac{\langle3p\rangle\langle p4\rangle}{[23][42]}+\frac{(1-\alpha't)}{\alpha'^{2}}\frac{\langle4p\rangle\langle p2\rangle}{[23][34]}\Big\},
\end{align}
\begin{align}
\mathscr{F}(j_z=0,-,-,-) & =C_{D_{2}}C\sqrt{2\alpha'}g^{3}\mathscr{T}_{n=2}^{a_{1}a_{2}a_{3}a_{4}}V_{t}(2\alpha')^{3}\frac{[qp]}{\sqrt{2}m}\times\Big\{\frac{(1-\alpha'u)}{\alpha'^{2}}\frac{\big(\langle3q\rangle\langle p2\rangle+\langle2q\rangle\langle p3\rangle\big)}{[34][42]}\nonumber \\
 & +\frac{(1-\alpha's)}{\alpha'^{2}}\frac{\big(\langle4q\rangle\langle p3\rangle+\langle3q\rangle\langle p4\rangle\big)}{[23][42]}+\frac{(1-\alpha't)}{\alpha'^{2}}\frac{\big(\langle4q\rangle\langle p2\rangle+\langle2q\rangle\langle p4\rangle\big)}{[23][34]}\Big\},
\end{align}
\begin{align}
\mathscr{F}(j_z=-1,-,-,-) & =C_{D_{2}}C\sqrt{2\alpha'}g^{3}\mathscr{T}_{n=2}^{a_{1}a_{2}a_{3}a_{4}}V_{t}(2\alpha')^{3}\frac{[pq]}{m}\times\Big\{\frac{(1-\alpha'u)}{\alpha'^{2}}\frac{\langle2q\rangle\langle q3\rangle}{[34][42]}\nonumber \\
 & \qquad+\frac{(1-\alpha's)}{\alpha'^{2}}\frac{\langle3q\rangle\langle q4\rangle}{[23][42]}+\frac{(1-\alpha't)}{\alpha'^{2}}\frac{\langle4q\rangle\langle q2\rangle}{[23][34]}\Big\}.
 \end{align}
and
\begin{align}
\mathscr{A}^{(4)}(\Xi_{1}^{\pm}(j_z),-,-,-) & =\big[\frac{5}{8}\mp(\frac{x_{1}}{6}-\frac{y_{1}}{3})
\big]\mathscr{F}(j_z,-,-,-),\\
\mathscr{A}^{(4)}(\Xi_{2}^{\pm}(j_z),-,-,-) & =\big[-\frac{5}{8}\mp(\frac{x_{2}}{6}-\frac{y_{2}}{3})
\big]\mathscr{F}(j_z,-,-,-).
\end{align}
Note that the constraints (\ref{const1},\ref{const2}) on the parameters $x,y$ automatically ensure the decoupling of $\Xi^+$ from all-minus configurations and of $\Xi^-$ from all-plus ones.

Next, we turn to  mostly plus configuration $(++-)$. {}For each $j_z=0,\pm 1$, there are two kinematic structures common to these amplitudes:
\begin{align}
\mathscr{K}_{1}(j_z=+1,+,+,-) & =C_{D_{2}}C\sqrt{2\alpha'}g^{3}\mathscr{T}_{n=2}^{a_{1}a_{2}a_{3}a_{4}}
V_{t}(2\alpha')^{3}\frac{\langle pq\rangle}{2m}\frac{[23]^{2}}{[24][34]}\frac{1-\alpha'u}{\alpha'}[2q][q3],\\
\mathscr{K}_{2}(j_z=+1,+,+,-) & =C_{D_{2}}C\sqrt{2\alpha'}g^{3}\mathscr{T}_{n=2}^{a_{1}a_{2}a_{3}a_{4}}V_{t}(2\alpha')^{3}\frac{[pq]}{2m}(-)\langle p4\rangle^{2}[23]^{2},
\end{align}
and
\begin{align}
\mathscr{K}_{1}(j_z=0,+,+,-) & =C_{D_{2}}C\sqrt{2\alpha'}g^{3}\mathscr{T}_{n=2}^{a_{1}a_{2}a_{3}a_{4}}V_{t}(2\alpha')^{3}\frac{1}{2\sqrt{2}}\frac{\langle pq\rangle}{m}\nonumber \\
 & \qquad\times\frac{[23]^{2}}{[24][34]}\frac{1-\alpha'u}{\alpha'}\big([2p][q3]+[3p][q2]\big),\\
\mathscr{K}_{2}(j_z=0,+,+,-) & =C_{D_{2}}C\sqrt{2\alpha'}g^{3}\mathscr{T}_{n=2}^{a_{1}a_{2}a_{3}a_{4}}V_{t}(2\alpha')^{3}\frac{1}{2\sqrt{2}}\frac{[pq]}{m}2\langle p4\rangle\langle q4\rangle[23]^{2},
\end{align}
and lastly
\begin{align}
\mathscr{K}_{1}(j_z=-1,+,+,-) & =C_{D_{2}}C\sqrt{2\alpha'}g^{3}\mathscr{T}_{n=2}^{a_{1}a_{2}a_{3}a_{4}}V_{t}(2\alpha')^{3}\frac{\langle pq\rangle}{2m}\frac{(-)[23]^{2}}{[24][34]}\frac{1-\alpha'u}{\alpha'}[2p][p3],\\
\mathscr{K}_{2}(j_z=-1,+,+,-) & =C_{D_{2}}C\sqrt{2\alpha'}g^{3}\mathscr{T}_{n=2}^{a_{1}a_{2}a_{3}a_{4}}V_{t}(2\alpha')^{3}\frac{[pq]}{2m}\langle q4\rangle^{2}[23]^{2}.
\end{align}
We obtain
\begin{align}
\mathscr{A}^{(4)}(\Xi_{1}^{\pm}(j_z),+,+,-) & =\big[\frac{5}{8}\pm(\frac{x_{1}}{6}-\frac{y_{1}}{3})\big]
\mathscr{K}_{1}(j_z)+\big[\frac{23}{16}\pm(\frac{7x_{1}}{12}+\frac{5y_{1}}{6})\big]
\mathscr{K}_{2}(j_z),\\
\mathscr{A}^{(4)}(\Xi_{2}^{\pm}(j_z),+,+,-) & =\big[-\frac{5}{8}\pm(\frac{x_{2}}{6}-\frac{y_{2}}{3})\big]\mathscr{K}_{1}(j_z)
+\big[\frac{13}{16}\pm(\frac{7x_{2}}{12}+\frac{5y_{2}}{6})\big]\mathscr{K}_{2}(j_z).
\end{align}
{}For the opposite, $(--+)$ helicity configurations,
\begin{align}
\mathscr{A}^{(4)}(\Xi_{1}^{\pm}(j_z),-,-.+) & =\big[\frac{5}{8}\mp(\frac{x_{1}}{6}-\frac{y_{1}}{3})\big]
\mathscr{K}_{1}(-j_z)^*+\big[\frac{23}{16}\mp(\frac{7x_{1}}{12}+
\frac{5y_{1}}{6})\big]\mathscr{K}_{2}(-j_z)^*,\\
\mathscr{A}^{(4)}(\Xi_{2}^{\pm}(j_z),-,-,+) & =\big[-\frac{5}{8}\mp(\frac{x_{2}}{6}-\frac{y_{2}}{3})\big]\mathscr{K}_{1}(-j_z)^*
+\big[\frac{13}{16}\mp(\frac{7x_{2}}{12}+\frac{5y_{2}}{6})\big]\mathscr{K}_{2}(-j_z)^*.
\end{align}
Note that the conditions (\ref{const1}) and (\ref{const2}) imply vanishing
$\mathscr{K}_{1}$ parts of  the $(\Xi^+, -, -, +)$  and $(\Xi^-, +, +, -)$ amplitudes.
By requiring that their $\mathscr{K}_{2}$ parts also vanish, we obtain
\begin{flalign}
\frac{23}{16}-(\frac{7}{12}x_{1}+\frac{5}{6}y_{1}) & =0,\\
\frac{13}{16}-(\frac{7}{12}x_{2}+\frac{5}{6}y_{2}) & =0,
\end{flalign}
which, combined with Eqs.(\ref{const1}) and (\ref{const2}) fixes the relative weights of vectors and pseudo-vectors to
\begin{equation}
\left\{ \begin{aligned} & x_{1}=3\\
 & y_{1}=-3/8
\end{aligned}\quad
\right.,\quad
\left\{ \begin{aligned} & x_{2}=-3/4\\
 & y_{2}=3/2
\end{aligned}
\right..\label{xxyy}
\end{equation}

To summarize, at the second massive level we identified two complex vectors, $\Xi_{1,2}$, with the vertex operators written in Eqs.(\ref{firstv}) and (\ref{secondv}) and the parameters $x$ and $y$
given in Eq.(\ref{xxyy}), which satisfy the following selection rules:
\begin{gather}
\mathscr{A}\left[\Xi^{+},-,-\right]=\mathscr{A}\left[\Xi^{\pm},+,-\right]=\mathscr{A}\left[\Xi^{-},+,+\right]=0,
\end{gather}
for three-point amplitudes and
\begin{gather}
\mathscr{A}\left[\Xi^{+},-,-,-\right]=\mathscr{A}\left[\Xi^{-},+,+,+\right]=0,\\
\mathscr{A}\left[\Xi^{+},+,-,-\right]=\mathscr{A}\left[\Xi^{+},-,+,-\right]=
\mathscr{A}\left[\Xi^{+},-,-,+\right]=0,\\
\mathscr{A}\left[\Xi^{-},-,+,+\right]=\mathscr{A}\left[\Xi^{-},+,-,+\right]=
\mathscr{A}\left[\Xi^{-},+,+,-\right]=0,
\end{gather}
for four-point amplitudes.The overall vertex normalization can be fixed by the usual factorization argument. It is
\begin{equation}
C=\frac{2}{5}\sqrt{\alpha'}g.
\end{equation}

\section{Factorization and BCFW Reconstruction of the Four-gluon Amplitude}

In this Section, we consider the $s$-channel residue expansion of the partial four-gluon MHV amplitude, $M(p\,-,q\,-,k_1\,+,k_2\,+)$, with the external momenta $p,~q, ~k_1,~k_2$ and the respective Chan-Paton factor $4g^2{\rm Tr}(T^{a_p}T^{a_q}T^{a_1}T^{a_2})$, with the coupling constant $g$ included. We want to compare the residues with the factorized sum\footnote{In this Section, we set the mass scale $M=1$.}
\begin{equation}\label{ffun}F(p\,-,q\,-,k_1\,+,k_2\,+)\equiv
\sum_{m_j,\,j<n}(p\,{-}, q\,{-} |m_j,j,n)\frac{1}{s-n}(k_2\,{-}, k_1\,{-}|m_j,j,n)^*\ ,
\end{equation}
where $s=2p\cdot q$ (we also define $u=2q\cdot k_1$ and $t=2q\cdot k_2$), and $(p\,{+}, q\,{+} |m_j,j,n)$ are the three-point on-shell amplitudes involving  two gluons and one string state at mass level $n$, with the spin quantum numbers $j,\,m_j$. The purpose of this exercise is to compare the three-point amplitudes with those evaluated in the previous Sections and to show explicitly how the four-gluon amplitude can be reconstructed by a BCFW deformation of the factorized sum:
\begin{equation}
F(p\,-,q\,-,k_1\,+,k_2\,+)\stackrel{\scriptscriptstyle BCFW}{\longrightarrow}M(p\,-,q\,-,k_1\,+,k_2\,+)\ .
\end{equation}

The four-gluon amplitude is given by
\begin{eqnarray}
M(p\,-,q\,-,k_1\,+,k_2\,+)&=&\frac{\langle pq\rangle^4}{\langle pq\rangle\langle q1\rangle\langle 12\rangle\langle 2p\rangle}\frac{\Gamma(1-s)\Gamma(1-u)}{\Gamma(1-s-u)}\nonumber\\[1mm]
&=&\langle pq\rangle^2[12]^2\,
\frac{1}{s}\,B(1-s,-u)
\end{eqnarray}
By using the well-known expansion of the Beta function:
\begin{equation}\label{mamp}
M(p\,-,q\,-,k_1\,+,k_2\,+)=\langle pq\rangle^2[12]^2\,
\frac{1}{s}\sum_{n=1}^{\infty}\frac{1}{n-s}\frac{(u+1)_{n-1}}{(n-1)!}\ ,
\end{equation}
The residue associated to the mass level $n$ is
\begin{equation}\label{resm}
\makebox{Res}_{s=n}M(p\,-,q\,-,k_1\,+,k_2\,+)=-\langle pq\rangle^2[12]^2\,\frac{(u+1)_{n-1}}{n!}\ .
\end{equation}
Note that the Pochhammer symbol contracts Lorentz indices across the $s$-channel (recall $u=2q\cdot k_1$).
The flow of Lorentz indices is due to the propagation of higher spin states in the $s$-channel. The first non-trivial contraction occurs at level $n=2$, where it is
due to massive vector particles discussed in the previous Section. At a given mass level $n$, not all spins $j$ propagate: only the even ones for odd $n$ and the odd ones for even $n$, up to $j=n-1$. For instance, at the next $n=3$ level, both $j=0$ and $j=2$ contribute. We want to compare Eq.(\ref{resm}) with the residues of the factorized sum (\ref{ffun}).

In the factorized sum (\ref{ffun}), two pairs of gluons, $(p,q)$ and $(k_1,k_2)$
are coupled through intermediate Regge particles propagating in the $s$-channel. The Lorentz indices are transferred by the wave functions of intermediate particles, depending on a fixed spin quantization axis defined by the choice of reference vectors. The most convenient spin quantization axis  is the direction of motion of the $(p,q)$ pair in its center of mass frame, which is imposed by choosing $p$ and $q$ as the reference vectors for the massive wave functions, see Appendix C.
In this case, the angular momentum conservation dictates that only $m_j=0$ states propagate in the factorized sum. Let us illustrate this point on the example of a massive vector particle.

In the previous Section, Eq.(\ref{3p--}), we found that, up to a numerical factor,
\begin{equation}
(p\,{-}, q\,{-} |m,j=1,n=2)
=\langle pq\rangle^2 (\xi_m \cdot q)\ .
\end{equation}
Indeed, with the choice of $(p,q)$ as the reference vectors for the polarization vectors $\xi_m$, one finds
 \begin{equation}
\xi_{{-}1} \cdot q\ =\xi_{{+}1} \cdot q\ =0\quad,\qquad \xi_{\, 0\,} \cdot q\ =\frac{1}{\sqrt{2}}\, (p-q)q =\frac{\sqrt{s}}{2}=\frac{1}{\sqrt{2}}~ ,
\end{equation}
where $\sqrt{2}=\sqrt{M_n}$ appear from the wave function normalization factors.
On the other hand, with the same choice of the reference vectors,
\begin{eqnarray}
(k_2\,{-}, k_1\,{-} |m=0,j=1,n=2)&=&\langle 12\rangle^2 \,\frac{1}{\sqrt{2}}\, (p-q)k_1\\ &=&-\langle 12\rangle^2 \,\frac{1}{\sqrt{2}}\,(u+s/2)=-\langle 12\rangle^2 \,\frac{1}{\sqrt{2}}\,(u+1)\ .\nonumber
\end{eqnarray}
In this way, we obtain
\begin{equation}
(p\,{-}, q\,{-} |0,1,2)(k_2\,{-}, k_1\,{-} |0,1,2)^*=-\langle pq\rangle^2[ 12]^2\frac{(u+1)}{2}\ ,
\end{equation}
in agreement with the residuum (\ref{resm}) for $n=2$.

It is clear that for the above choice of reference vectors, the residues at $s=n$ of the factorized sum have the form $\langle pq\rangle^2[ 12]^2$ times a function of
\begin{equation}
a\equiv (p-q)k_2=-(p-q)k_1=u+\frac{s}{2}=u+\frac{n}{2}\ ,
\end{equation}
where the last step follows from the on-shell condition for the massive particle.
We can obtain the factorized sum by simply setting $u=a-\frac{n}{2}$
in Eq.(\ref{mamp})
\begin{eqnarray}\nonumber
F(p\,-,q\,-,k_1\,+,k_2\,+)&=&\langle pq\rangle^2[12]^2\sum_{n=1}^{\infty}\frac{\makebox{Res}_{(s=n\, , ~u=a-\frac{n}{2})}M(p\,-,q\,-,k_1\,+,k_2\,+)}{s-n}\\ &=&
\langle pq\rangle^2[12]^2
\sum_{n=1}^{\infty}\frac{1}{n-s}\frac{(a-\frac{n}{2}+1)_{n-1}}{n!}\ .\label{fresult}\end{eqnarray}
We checked the above result also at the $n=4$ level, by combining the
on-shell amplitudes involving spin $j=0$ and $j=2$ Regge states, according to Eq.(\ref{ffun}). It is convenient to introduce the generating function
\begin{equation}\label{gfun}
g_F(x)=\sum_{n=1}^{\infty}\frac{(a-\frac{n}{2}+1)_{n-1}}{n!}\, x^{n-1}\ ,
\end{equation}
so that
\begin{equation}
F(p\,-,q\,-,k_1\,+,k_2\,+)=\langle pq\rangle^2[12]^2\int_{0}^1 x^{-s}\,g_F(x)\ .
\end{equation}
It is easy to see that the generating function (\ref{gfun}) satisfies
\begin{equation}
\frac{d}{dx}\big[ x g_F(x)\big]=(1+{\textstyle \frac{x^2}{4}})^{-1/2}\, e^{2a{ \rm ArcSinh}(\frac{x}{2})}\ .
\end{equation}
We want to stress again the the factorized sum is evaluated by using on-shell amplitudes involving one massive state and two gluons. We will show how to reconstruct the four gluon amplitude of Eq.(\ref{mamp}), which involves intermediate particles propagating off-shell, by applying a BCFW deformation to $F$, Eq.(\ref{fresult}).

It has been argued recently that the BCFW recursion relations, originally formulated for pure Yang-Mills theory \cite{CSW}-\cite{LecBCFW}, hold also in string theory \cite{StrBCFW1}-\cite{Fot3}. The arguments rely crucially on proving the absence of an essential singularity at $z\to\infty$ ($z$ is the deformation parameter) of the full-fledged string amplitudes.  The proof is straightforward for four-gluon amplitudes but becomes increasingly complex for more gluons. Note that in the string case, there is an infinite number of intermediate states propagating in any channel, as seen explicitly in the factorized sum (\ref{ffun}). This should be contrasted with the Yang-Mills case \cite{PTMHV}, where there are no massless on-shell states propagating in the $s$-channel of the deformed $(-\,-\,+\,+)$ amplitude.

In order to force the $s$-channel resonances on-shell, we apply the BCFW deformation
\begin{equation}\label{deform}
p\to \hat{p}=p-zv~,\qquad k_2\to \hat{k_2}=k_2+zv\ ,
\end{equation}
where the light-like vector
\begin{equation}
v^\mu=\langle p|\sigma^\mu |2]
\end{equation}
and $z$ is the deformation parameter. Since $\hat{s}=s-2z\,vq$, the resonance poles appear at
\begin{equation}
z=\frac{s-n}{2vq}\ .
\end{equation}
Under this deformation
\begin{equation}
a=(p-q)k_2 ~\to~ \hat{a}=a-z\, vq=a-\frac{s}{2}+\frac{n}{2}=u+\frac{n}{2}\ .
\end{equation}
Upon $a\to\hat{a}$, the generating function (\ref{gfun}) transforms into
\begin{equation}\label{gmfun}
g_F(x)\stackrel{\scriptscriptstyle BCFW}{\longrightarrow} g_M(x)
=\sum_{n=1}^{\infty}\frac{(u+1)_{n-1}}{n!}\, x^{n-1}\ ,
\end{equation}
which satisfies
\begin{equation}\label{parts}
\frac{d}{dx}\big[ x g_M(x)\big]=(1-x)^{-u-1}\ .
\end{equation}
In this way, we obtain
\begin{eqnarray}\nonumber
M(p\,-,q\,-,k_1\,+,k_2\,+) &=& \langle pq\rangle^2[12]^2\int_{0}^1 x^{-s}\,g_M(x)\\
&=&
\langle pq\rangle^2[12]^2\,\frac{1}{s}\,B(1-s,u)\ ,
\end{eqnarray}
where we used Eq.(\ref{parts}) to integrate by parts. As usual with world-sheet duality, it is rewarding to see how the massless gluon pole appears in the $u$-channel after summing over the $s$-channel exchanges of massive string states.

Apart from providing the first explicit example of a BCFW construction in string theory, the above example seems of little or no practical importance. After all, what more can we learn by dissecting the Veneziano-Virasoro-Shapiro amplitude?
It would be interesting, however, to construct all multi-gluon string disk amplitudes by using the BCFW recursion relations.  Unfortunately, it is not so easy: starting from five gluons, a standard BCFW deformation, like in Eq.(\ref{deform}), yields on-shell poles in two channels, and the step leading from the factorized sums to the actual amplitude becomes quite cumbersome. Even in bosonic string theory, it is not clear how to combine much simpler factorized sums with five external tachyons to the well known five-tachyon amplitude.\\[5mm]
\textbf{Acknowledgements}\\[5mm] WZ.F.\ is grateful
to HaiPeng An, Ning Chen, LiPeng Lai, Oliver Schlotterer, ChaoLun Wu and
Peng Zhou for helpful discussions. T.R.T.\ is grateful to Rutger Boels,
Niels Obers and Brian Wecht for useful correspondence.
This material is based in part upon work supported by the National Science Foundation under Grant No.\ PHY-0757959.  Any
opinions, findings, and conclusions or recommendations expressed in
this material are those of the authors and do not necessarily reflect
the views of the National Science Foundation.

\section*{Appendix}

\appendix

\section{Wigner d-matrix}

The Wigner D-matrix (a.k.a. Wigner rotation matrix), introduced in
1927 by Eugene Wigner, is a dimension $2j+1$ square matrix, which
is in an irreducible representation of groups $SU(2)$ and $SO(3)$.
The matrix is defined to be:
\begin{equation}
D_{m',m}^{(j)}(\alpha,\beta,\gamma)=\langle jm'|R(\alpha,\beta,\gamma)|jm\rangle=e^{-im'\alpha}d_{m',m}^{(j)}(\beta)e^{-im\gamma},
\end{equation}
where $\alpha,\beta,\gamma$ are Euler angles, and $d_{m',m}^{j}(\beta)$,
known as Wigner reduced (or small) $d-$matrix, is given by a general
formula \cite{Wigner,Edmonds}:

\begin{align}
d_{m',m}^{(j)}(\beta) & =\sqrt{\frac{(j+m')!(j-m')!}{(j+m)!(j-m)!}}\sum_{s}(-)^{j-m'-s}\nonumber \\
 & \qquad\times\left(\begin{array}{c}
j+m\\
j-m'-s
\end{array}\right)\left(\begin{array}{c}
j-m\\
s
\end{array}\right)\left(\cos\frac{\beta}{2}\right)^{m'+m+2s}\left(\sin\frac{\beta}{2}\right)^{2j-m'-m-2s}.\label{W-dm}
\end{align}
The sum over $s$ is over such values that the factorials are non
negative.
Two important relations follow from the
above expression:
\begin{equation}
d_{0,0}^{(l)}(\theta)=P_{l}(\cos\theta),
\end{equation}
where $P_{l}(cos\theta)$ is the Legendre polynomial, and
\begin{equation}
d_{m',m}^{(j)}(\theta)=(-1)^{j-m}d_{m',-m}^{(j)}(\theta+\pi).\label{inversem}
\end{equation}

{}For $j\leq 4$, the following Wigner d-matrices appear in the factorized four-gluon amplitudes:
\begin{itemize}
\item $d_{0,0}^{(j)}(\theta)$
\end{itemize}
\begin{align}
d_{0,0}^{(0)}(\theta) & =P_{0}(\cos\theta)=1,\\
d_{0,0}^{(1)}(\theta) & =P_{1}(\cos\theta)=\cos\theta,\\
d_{0,0}^{(2)}(\theta) & =P_{2}(\cos\theta)=\frac{1}{2}(3\cos^{2}\theta-1),\\
d_{0,0}^{(3)}(\theta) & =P_{3}(\cos\theta)=\frac{1}{2}(5\cos^{3}\theta-3\cos\theta),\\
d_{0,0}^{(4)}(\theta) & =P_{4}(\cos\theta)=\frac{1}{8}(35\cos^{4}\theta-30\cos^{2}\theta+3).
\end{align}

\begin{itemize}
\item $d_{2,\pm 2}^{(j)}(\theta)$
\end{itemize}
\begin{align}
d_{2,\pm 2}^{(2)}(\theta) & =\left(\frac{1\pm\cos\theta}{2}\right)^{2},\\
d_{2,\pm 2}^{(3)}(\theta) & =\frac{1}{4}(3\cos^{3}\theta\pm 4\cos^{2}\theta-\cos\theta\mp 2),\\
d_{2,\pm  2}^{(4)}(\theta) & =\frac{1}{4}(7\cos^{4}\theta\pm 7\cos^{3}\theta-6\cos^{2}\theta\mp 5\cos\theta+1).
\end{align}

\begin{itemize}
\item $d_{2,\pm1}^{(j)}(\theta)$
\end{itemize}
\begin{align}
d_{2,\pm 1}^{(2)}(\theta) & =\frac{1}{2}\sin\theta(1\pm \cos\theta),\\
d_{2,\pm 1}^{(3)}(\theta) & =\frac{\sqrt{10}}{8}\sin\theta(\pm 3\cos^{2}\theta+2\cos\theta\mp 1),\\
d_{2,\pm 1}^{(4)}(\theta) & =\frac{\sqrt{2}}{8}\sin\theta(\pm 14\cos^{3}\theta+7\cos^{2}\theta\mp 8\cos\theta-1).
\end{align}

\section{Computation of $c_k^{(n)}$ coefficients}
In the limit of $s\rightarrow nM^{2}$, $V_{s}$ is regular while both
$V_{t}$ and $V_{u}$ have single poles.
By using (\ref{Bexpansion}) and (\ref{Vexpansion}), we
find,
\begin{equation}
{\rm Res}_{s=nM^2}V_{t}/u  =(-1)^n{\rm Res}_{s=nM^2}V_{u}/t\label{Vtuevenn}
\end{equation}
Thus, for the purpose of extracting the residues at $s=nM^2$, we can use $V_{t}=u\times V(\cos\theta,n)$,
and $V_{u}=(-1)^n t\times V(\cos\theta,n)$, where
$V(\cos\theta,n)$ is a common factor.
 Thus, for odd $n$, the MHV amplitude
(\ref{MHV--++}) can be written as
\begin{align}
\mathcal{M}(g_{1}^{-},g_{2}^{-},g_{3}^{+},g_{4}^{+}) & \rightarrow4g^{2}\langle12\rangle^{4}
\Big[\frac{u V
(\cos\theta,n)}{\langle12\rangle\langle23\rangle\langle34\rangle\langle41\rangle}
Tr(T^{a_{1}}T^{a_{2}}T^{a_{3}}T^{a_{4}}+T^{a_{2}}T^{a_{1}}T^{a_{4}}T^{a_{3}})
\nonumber\\
 & \qquad\qquad\;\,+\frac{t V(\cos\theta,n)}{\langle13\rangle\langle34\rangle
 \langle42\rangle\langle21\rangle}
 Tr(T^{a_{2}}T^{a_{1}}T^{a_{3}}T^{a_{4}}+T^{a_{1}}T^{a_{2}}T^{a_{4}}T^{a_{3}})\Big]
 \nonumber\\
 & \!\!\!\!\!  \!\!\!\!\!=\frac{\langle12\rangle^{4}u V(\cos\theta,n)}{\langle12\rangle\langle23
 \rangle\langle34\rangle\langle41\rangle}Tr(\{T^{a_{1}},T^{a_{2}}\}\{T^{a_{3}},T^{a_{4}}\})
 =S^{a_1a_2}_{a_3a_4}\frac{\langle12\rangle^{4}V_{t}(\cos\theta,n)}{\langle12\rangle\langle23\rangle\langle34\rangle\langle41\rangle}\nonumber\\
 & \sim S^{a_1a_2}_{a_3a_4}\frac{s}{u}V_{t}(\cos\theta,n)\qquad\makebox{(up to a phase factor)}.\nonumber
\end{align}
{}For even $n$, $S^{a_1a_2}_{a_3a_4}\to A^{a_1a_2}_{a_3a_4}$.
To summarize, the MHV amplitude (\ref{MHV--++}) can be rewritten
as,
\begin{align}
\mathcal{M}(g_{1}^{-},g_{2}^{-},g_{3}^{+},g_{4}^{+}) & \rightarrow S^{a_1a_2}_{a_3a_4}\frac{s}{u}V_{t}(\cos\theta,n)\qquad(n\; odd),\\
\mathcal{M}(g_{1}^{-},g_{2}^{-},g_{3}^{+},g_{4}^{+}) & \rightarrow A^{a_1a_2}_{a_3a_4}\frac{s}{u}V_{t}(\cos\theta,n)\qquad(n\; even).
\end{align}
The function $\frac{s}{u}V_{t}(\cos\theta ,n)\equiv F(\cos\theta,n)$ is a polynomial
of $\cos\theta$, and can be decomposed into the sum of
Legendre polynomials $P_l(x\equiv\cos\theta)=d_{0,0}^{(l)}(\theta)$.
We will see that for even $n$, $F(x,n)$ contains
odd powers of $x$ only, ranging from  $1$ up to  $n-1$; for odd $n$,
$F(x,n)$ contains even powers only, ranging from $0$ up to
$n-1$. Thus for even $n$, $F(x,n)$ can
be written as  a sum of odd Legendre polynomials while for
odd $n$ it be written as a sum of even Legendre
polynomials. Recall
\begin{equation}
F(x,n)\equiv\frac{s}{u}V_{t}(n)=\frac{nM^{2}}{s-nM^{2}}
\times\frac{M^{2-2n}}{(n-1)!}\prod_{J=1}^{n-1}(u+M^{2}J).
\end{equation}

\subsection{$n$ odd}

We write
\begin{equation}
F(x,n)\rightarrow\frac{M^{2}}{s-nM^{2}}f(x,n),
\end{equation}
where
\begin{align}
f(x,n) & =\Big[\frac{n}{(n-1)!}\Big]\underbrace{[\frac{nx}{2}-(\frac{n}{2}-1)][\frac{nx}{2}-(\frac{n}{2}-2)]
\cdots[\frac{nx}{2}+(\frac{n}{2}-2)][\frac{nx}{2}+(\frac{n}{2}-1)]}_{n-1\; factors}\nonumber \\
 & =\Big[\frac{n}{(n-1)!}\Big]\left(\frac{nx}{2}-\frac{n}{2}+1\right)_{(n-1)}.
\end{align}
The Pochhammer Symbol,
\begin{equation}
(x)_{n}=\frac{\Gamma(x+n)}{\Gamma(x)}=x(x+1)...(x+n-1)=\sum_{k=0}^{n}(-1)^{n-k}s(n,k)x^{k},\label{Stirling1}
\end{equation}
where $s(n,k)$ is the Stirling number of the first kind.
We obtain
\begin{align}
f(x,n) & =\Big[\frac{n}{(n-1)!}\Big]\left(\frac{nx}{2}-\frac{n}{2}+1\right)_{(n-1)}\nonumber \\
 & =\frac{n}{(n-1)!}(a_{n-1}^{(n)}x^{n-1}+a_{n-3}^{(n)}x^{n-3}+\cdots+a_{2}^{(n)}x^{2}+a_{0}^{(n)}).
\end{align}
with
\begin{equation}
a_{k}=\sum_{i=0}^{n-1-k}\frac{(-1)^{n-1-k}}{2^{k+i}}s(n-1,k+i)\left(\begin{array}{c}
k+i\\
i
\end{array}\right)(n)^{k}(n-2)^{i}.
\end{equation}
We want to rewrite $f(x,n)$ as
\begin{align}
f(x,n) & =\frac{n}{(n-1)!}(a_{n-1}^{(n)}x^{n-1}+a_{n-3}^{(n)}x^{n-3}+\cdots+a_{2}^{(n)}x^{2}+a_{0}^{(n)})\nonumber \\
 & =c_{n-1}^{(n)}P_{n-1}+c_{n-3}^{(n)}P_{n-3}+\cdots+c_{2}^{(n)}P_{2}+c_{0}^{(n)}P_{0}.
\end{align}
By using the formula,
\begin{equation}
x^{n}=\sum_{i=n,n-2,\cdots}\frac{(2i+1)n!}{2^{(n-i)/2}(\frac{n-i}{2})!(n+i+1)!!}P_{i}(x),
\label{pform}\end{equation}
we obtain
\begin{align}
c_{k}^{(n)} & =\frac{n}{(n-1)!}\sum_{j=0}^{\frac{n-1-k}{2}}a_{k+2j}d_{k+2j}\nonumber \\
 & =\frac{n}{(n-1)!}\sum_{j=0}^{\frac{n-1-k}{2}}\sum_{i=0}^{(n-1-k-2j)}\frac{(-1)^{n-1-k-2j}}{2^{k+3j+i}}\frac{(2k+1)(k+2j)!}{j!(2k+2j+1)!!}\nonumber \\
 & \qquad\qquad\qquad\qquad\times\left(\begin{array}{c}
k+2j+i\\
i
\end{array}\right)(n)^{k+2j}(n-2)^{i}s(n-1,k+2j+i).
\end{align}

The first coefficient is fairly easy to calculate, and it reads
\begin{equation}
c_{n-1}^{(n)}=\frac{n^{n}(n-1)!}{(2n-2)!}.\label{1stcoeff}
\end{equation}

\subsection{$n$ even}

For even $n$, the expansion of $f(x,n)$ is slightly different, but the analysis is similar. We have
\begin{align}
f(x,n) & =\Big[\frac{n}{(n-1)!}\Big](-1)\underbrace{[\frac{n}{2}x-(\frac{n}{2}-1)][\frac{n}{2}x-(\frac{n}{2}-2)]\cdots[\frac{n}{2}x+(\frac{n}{2}-2)][\frac{n}{2}x+(\frac{n}{2}-1)]}_{n-1\; factors}\nonumber \\
 & =\Big[\frac{n}{(n-1)!}\Big](-1)\left(\frac{nx}{2}-\frac{n}{2}+1\right)_{(n-1)}.
\end{align}
Here again, we write $f(x,n)$ as
\begin{align}
f(x,n) & =\Big[\frac{n}{(n-1)!}\Big](-1)\left(\frac{nx}{2}-\frac{n}{2}+1\right)_{(n-1)}\nonumber \\
 & =\frac{(-1)n}{(n-1)!}(a_{n-1}^{(n)}x^{n-1}+a_{n-3}^{(n)}x^{n-3}+\cdots+a_{3}^{(n)}x^{3}+a_{1}^{(n)}x).
\end{align}with
\begin{equation}
a_{k}=\sum_{i=0}^{n-1-k}\frac{(-1)^{n-1-k}}{2^{k+i}}s(n-1,k+i)\left(\begin{array}{c}
k+i\\
i
\end{array}\right)(n)^{k}(n-2)^{i}.
\end{equation}
We want to rewrite  $f(x,n)$ as
\begin{align}
f(x,n) & =\frac{n^{n}}{(n-1)!}(a_{n-1}^{(n)}x^{n-1}+a_{n-3}^{(n)}x^{n-3}+\cdots+a_{3}^{(n)}x^{3}+a_{1}^{(n)}x)\nonumber \\
 & =c_{n-1}^{(n)}P_{n-1}+c_{n-3}^{(n)}P_{n-3}+\cdots+c_{3}^{(n)}P_{3}+c_{1}^{(n)}P_{1}.
\end{align}
By using the formula (\ref{pform}), we obtain
\begin{align}
c_{k}^{(n)} & =\frac{(-1)n}{(n-1)!}\sum_{j=0}^{\frac{n-1-k}{2}}a_{k+2j}d_{k+2j}\nonumber \\
 & =\frac{(-1)n}{(n-1)!}\sum_{j=0}^{\frac{n-1-k}{2}}\sum_{i=0}^{(n-1-k-2j)}\frac{(-1)^{n-1-k-2j}}{2^{k+3j+i}}\frac{(2k+1)(k+2j)!}{j!(2k+2j+1)!!}\nonumber \\
 & \qquad\qquad\qquad\qquad\times\left(\begin{array}{c}
k+2j+i\\
i
\end{array}\right)(n)^{k+2j}(n-2)^{i}s(n-1,k+2j+i).
\end{align}

\section{Massive spin 1 Wave Functions}

A spin $j$ particle contains $2j+1$ spin degrees of freedom associated
to the eigenstates of $j_{z}$. The choice of the quantization axis
$z$ can be handled in an elegant way by decomposing the momentum
k into two arbitrary light-like reference momenta $p$ and $q$:
\begin{equation}
k^{\mu}=p^{\mu}+q^{\mu},\qquad k^{2}=m^{2}=2pq,\qquad p^{2}=q^{2}=0.
\end{equation}
Then the spin quantization axis is chosen as the direction of $q$
in the rest frame. The $2j+1$ spin wave functions depend of $p$
and $q$, however this dependence drops out in the amplitudes summed
over all spin directions and in \textquotedblleft{}unpolarized\textquotedblright{}
cross sections.
The massive spin-1 wave functions $\xi_{\mu}$ (transverse, i.e.,
$\xi_{\mu}k^{\mu}=0$) are given by the following polarization vectors
\cite{HWF1,HWF3} (up to a phase factor):
\begin{flalign}
\text{\ensuremath{\xi}}_{+}^{\mu}(k) & =\frac{1}{\sqrt{2}m}p_{\dot{a}}^{*}\bar{\sigma}^{\mu\dot{a}a}q_{a},\\
\text{\ensuremath{\xi}}_{0}^{\mu}(k) & =\frac{1}{2m}\bar{\sigma}^{\mu\dot{a}a}(p_{\dot{a}}^{*}p_{a}-q_{\dot{a}}^{*}q_{a}),\\
\text{\ensuremath{\xi}}_{-}^{\mu}(k) & =-\frac{1}{\sqrt{2}m}q_{\dot{a}}^{*}\bar{\sigma}^{\mu\dot{a}a}p_{a}.
\end{flalign}

\end{document}